\documentclass[aps,prl,preprint,superscriptaddress]{revtex4-2}
\usepackage{graphicx}
\graphicspath{ {./} }
\usepackage{amsmath}

\begin{document}

\title{Tunable piezoelectric metamaterial for Lamb waves using periodic shunted circuits}

\author{David R. Schipf}
\affiliation{NRC Research Associate Program, U.S. Naval Research Laboratory, Code 7160, Washington, D.C. 20375, USA}

\author{Matthew D. Guild}
\email{matthew.guild@nrl.navy.mil}
\affiliation{U.S. Naval Research Laboratory, Code 7160, Washington, D.C. 20375, USA}

\author{Caleb F. Sieck}
\affiliation{U.S. Naval Research Laboratory, Code 7160, Washington, D.C. 20375, USA}

\begin{abstract}
Piezoelectric elastic metamaterials offer the ability to overcome the fixed, narrow bandwidth characteristics of passive elastic metamaterials. Interesting ultrasonic band gaps exist in piezoelectric plate metamaterials with periodic electrodes connected to shunted circuits. These band gaps result from an avoided crossing between electrical and mechanical bands, and can arise at lower frequencies than Bloch wave band gaps. Current analytical modeling techniques for these systems are numerically cumbersome, and assume an infinitely periodic plate. We present an approximate two-dimensional analytical model that can be used to directly calculate scattering coefficients for finite length plates. This model is shown to predict a band diagram that compares well with diagrams obtained from finite element analysis (FEA). Lower than 10\% difference in the estimation of the location of the band gap was found for a plate thickness of $2$ mm, electrode width of $1$ mm, and gap between electrodes greater than $1.2$ mm. We calculate effective impedances and effective wavenumbers from global scattering coefficients. The calculated effective normalized wavenumber swings from positive values ($0<k_{\mathrm{eff}}\leq 1$) to negative values ($0>k_{\mathrm{eff}}\geq -1$) at the low-frequency band gap, resembling wavenumbers for negative stiffness Helmholtz resonator metamaterials. This presents a new perspective on periodic shunted circuit piezoelectric plates as electrically tunable, negative stiffness metamaterials analogous to Helmholtz resonator lined acoustic waveguides.
\end{abstract}

\maketitle

\section{Introduction}
Elastic metamaterials are engineered structures containing deeply sub-wavelength resonant elements that enable extreme, macroscopic effective properties beyond the bounds of traditional composite materials \cite{haberman2016acoustic}.  The effective properties obtained with metamaterials include extremely large, near-zero and negative material properties, such as mass density and bulk modulus \cite{haberman2016acoustic,cummer2016controlling}.  This leads to a wide range of interesting wave propagation phenomena, including near-zero and negative refractive indices, and artificial band gaps \cite{haberman2016acoustic,cummer2016controlling}. For passive metamaterials and phononic crystals, the frequency and bandgaps over which these phenomena occur is fixed for a given geometry.  Active metamaterials, including those using piezoelectric materials referred to as piezoelectric metamaterials, offer the ability to create artificial band gaps at arbitrarily low frequencies within the homogenization limit that can be conveniently tuned in the electrical domain. Tunable band gaps are an attractive capability for nano-scale energy transport \cite{cha2018electrical}, selective wave filtering \cite{trainiti2019time}, and vibration isolation \cite{wang2016tunable}. Rapid and stable tuning of band gaps enables non-reciprocal elastic wave propagation from space-time modulation \cite{croenne2019non,croenne2017brillouin,marconi2020experimental}. The electro-mechanical coupling found in piezoelectric materials enables a change in electrical circuit elements, connected to conductive boundaries, to change the elastodynamics within the material \cite{croenne2019non,croenne2017brillouin,marconi2020experimental}. 

The dispersion relation of a piezoelectric plate with periodically spaced electrical shunted circuits on the top and bottom was recently investigated theoretically\cite{kherraz2018hybridization,kherraz2019tunable} and experimentally \cite{chikh2019piezoelectric}. These investigations found relatively low frequency band gaps in the zeroth order symmetric Lamb wave $S_0$ resulting from the avoided crossing of the electrical circuit lattice band with the mechanical lamb wave band \cite{kherraz2018hybridization,kherraz2019tunable,chikh2019piezoelectric}. These \emph{hybridization} band gaps occur at lower frequencies than the Bloch-wave band gap arising from the periodicity of the electrodes connected to the circuits \cite{kherraz2019tunable,kherraz2018hybridization}. Due to the connection of these band gaps to the complex impedance of the shunted circuits, these band gaps can be tuned with variable electrical elements such as resistors, capacitors, or inductors. 

The purpose of this article is to further study these tunable hybridization band gaps with a pseudo-analytical modeling technique that allows direct calculation of scattering parameters for finite length plates. We use a two-dimensional (2D) model to divide a piezoelectric plate into a one-dimensional (1D) network of coupled cells, each with homogeneous properties and boundary conditions. The presented model yields an impedance matrix for each cell, relating the velocity and electric current to force and electric potential. Band diagrams calculated with this method are compared to band diagrams calculated from finite element analysis (FEA) of a single unit cell. Our findings further elucidate the hybridization band gap findings in \cite{kherraz2018hybridization,kherraz2019tunable}, and our model gives researchers a method for calculating scattering parameters and effective parameters for finite length piezoelectric plates with individually tunable shunted circuits.

\section{Background}
Analytical modeling of a piezoelectric plate with periodically spaced shunted circuits and electrodes was previously done using plane wave expansion (PWE), and the full set of piezoelectric equations for a bulk ceramic \cite{kherraz2019tunable}. While an analytical model for numerical calculation of phase velocities within a piezoelectric plate with homogeneous boundary conditions has been available for some time (\cite{joshi1991propagation}), the Kherraz et al. 2019 study \cite{kherraz2019tunable} recently modeled the problem with periodic electrodes forming an infinite phononic crystal. 

The model for the study in \cite{kherraz2019tunable}, and the study in this article, is a piezoelectric plate of thickness $h$, covered periodically with strip electrodes on the top and the bottom. As shown in Figure \ref{fig:first_diagram}, the width of the electrodes is $a_1$, while the width of the uncovered segments is $a_2$. The bottom electrodes can be connected to a shunted circuit with a complex impedance, as was the case for some of the studies in \cite{kherraz2019tunable}, or it can be grounded. The top electrodes are each connected to a shunted circuit with a complex impedance $Z_\mathrm{L}$. 

\begin{figure}[!ht]
\includegraphics{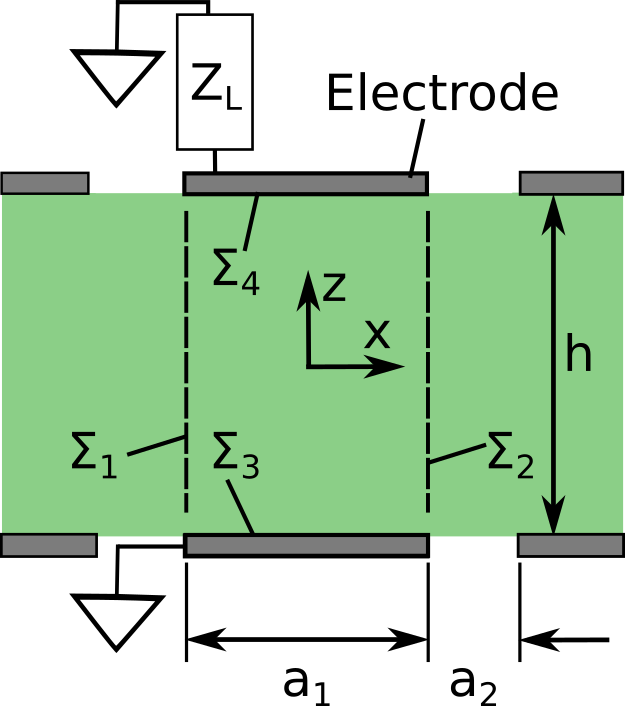}%
\caption{\label{fig:first_diagram}: Cross-sectional diagram of a piezoelectric plate, covered with periodically spaced electrodes, and loaded with shunted circuits. The dotted lines segment a portion of the plate covered by electrodes with homogeneous boundary conditions.}
\label{fig:first_diagram}
\end{figure}

The governing equations for macroscopic electro-mechanical behavior in piezoelectric plates can be written with indicial notation as \cite{wilson1988introduction}
\begin{equation}
T_{ij}=c^\mathrm{E}_{ijkl}\epsilon_{kl}-e_{kij}E_k ,
\label{eq:piezo_1}
\end{equation}
\begin{equation}
D_i=e_{ikl}\epsilon_{kl}+\varepsilon^\mathrm{S}_{ik}E_k ,
\label{eq:piezo_2}
\end{equation}
where $T_{ij}$ is the stress tensor, $\epsilon_{kl}$ is the strain tensor, $D_i$ is the dielectric displacement vector, $E_k$ is the electric field vector, $c^\mathrm{E}_{ijkl}$ is the stiffness tensor, $e_{kij}$ is the piezoelectric coefficient tensor, and $\varepsilon^\mathrm{S}_{ik}$ is the permittivity tensor. The continuity equation for stress propagation in a solid is
\begin{equation}
T_{ij,i}=\rho \ddot{\xi}_j ,
\label{eq:continuity}
\end{equation}
where $\rho$ is the density, and ${\xi}_j$ are the displacement components. The electric field components can be related to the electric potential $\phi$ with
\begin{equation}
E_k=-\phi_{,k} .
\label{eq:electric_field}
\end{equation}
Strain can be related to displacements by
\begin{equation}
\epsilon_{ij} = \frac{1}{2}(u_{i,j}+u_{j,i}) ,
\label{eq:strain}
\end{equation}
and the dielectric displacement within the plate must satisfy
\begin{equation}
D_{i,i}=0 ,
\label{eq:dielectric_displacement}
\end{equation}
due to a lack of free charge.

The solution assumed in \cite{kherraz2019tunable} for the displacement within the solid is
\begin{equation}
\xi_j=\sum_{n=-\infty}^{+\infty} e^{-i\omega s_{1n} x}\sum_{p=1}^6C_n^{p}G_{jn}^{p}e^{-i\omega s_{3n}^p z} ,
\label{eq:assumed_displacement}
\end{equation}
with $j=1,2,3$ corresponding to the displacements in the $x$,$y$, and $z$ directions, respectively. The slowness vectors for the $n$th Fourier component in the $x$ direction and the $z$ direction are $s_{1n}$ and $s_{3n}$, respectively. The coefficients for the $n$th Fourier component are given as $C_n^{p}$ and $G_{jn}^{p}$. The $n$th wavenumber is $k_{1n}=k_1+2\pi/(a_1+a_2)$. Given the above solution for the displacements, a solution for the electric potential within the solid is necessitated to be
\begin{equation}
\phi=\sum_{n=-\infty}^{+\infty} e^{-i\omega s_{1n} x}\sum_{p=1}^6C_n^{p}G_{4n}^{p}e^{-i\omega s_{3n}^p z} .
\label{eq:assumed_potential}
\end{equation}
The six polarization components $p=1,2,\dots,6$ come from an eigenvalue equation with six eigenvalue solutions $s_{3n}^p$. The bulk piezoelectric equations (\ref{eq:piezo_1})--(\ref{eq:dielectric_displacement}) can be re-arranged into a symmetric $4\times4$ matrix $\mathbf{M}$ that is multiplied by the coefficient vector for the three displacements and electric potential $\mathbf{G}=[G_1,G_3,G_4,G_2]$. The eigenvalue equation, given a value for the $n$th slowness vector component $s_{1n}$, is \cite{kherraz2019tunable}  

\begin{equation}
\mathbf{M}(s_{3n})\mathbf{G}=\mathbf{0} .
\end{equation}

In \cite{kherraz2019tunable}, the mechanical and electrical boundary conditions are formulated into an additional eigenvalue equation $\mathbf{Q}(\omega)\mathbf{C}=\mathbf{0}$ of size $4\times(2n_{max}+1)$, where $n_{max}$ is the number of Fourier coefficients used. To find the eigenfrequencies $\omega$ from a defined value of $k_1$ requires an iterative process and the numerical solution of two eigenvalue matrix problems. Band diagrams constructed with this model matched band diagrams constructed using FEA for grounded and floating electrodes on the top and bottom of the plate \cite{kherraz2019tunable}.

While this analytical model can produce band diagrams for infinitely periodic plates, it does not directly produce scattering parameters for a plate of finite length. An alternative approach to modeling this problem is to construct a set of equations for each segment of the plate that has homogeneous properties and boundary conditions. We present this approach, and detail how each segment can be coupled to form a global scattering matrix for an entire plate of finite length.

\section{Approximate Analytical Model}
\label{sec:methods}
An analytical model for symmetric $0$th order Lamb waves (acoustic waves) in piezoelectric plate segments with electrodes connected to shunted circuits, and plate segments without electrodes, is presented. The final form of the model is an impedance matrix equation for each discrete segment of the plate, herein called cells, with homogeneous boundary conditions, thickness, and properties. A method for calculating band diagrams of infinitely periodic plates based on the scattering matrix of each cell is given. A method for coupling the scattering matrices of each cell to formulate a global scattering matrix for a plate containing multiple cells is also given. Finally, equations for effective impedance and effective wavenumber, calculated from scattering parameters, is detailed.
 
\subsection{Impedance Matrix Formulation}
For this study, the $y$ direction width of the piezoelectric plate is assumed to be $w \gg \Lambda$, where $\Lambda$ is the width of one unit-cell, consisting of one open (uncovered) cell, and one electroded (covered) cell $\Lambda=a_1+a_2$. We assume that the thickness $h$ of the plate is significantly less than the width and the length of the entire plate $h\ll w$, $h\ll N\Lambda$, where $N$ is the number of unit-cells. We assume that the electrode thickness $T$ is much smaller than the plate thickness $T/h \ll 1$, as was the case in previous piezoelectric plate studies \cite{kherraz2018hybridization,kherraz2019tunable}. 

We further narrow this study to certain piezoceramics with non-zero piezoelectric coefficients $e_{31}$, $e_{33}$, $e_{32}$, $e_{24}$, $e_{15}$, and permittivity $\varepsilon^\mathrm{S}_{11}=\varepsilon^\mathrm{S}_{22}\neq\varepsilon^\mathrm{S}_{33}$, which comes from poling in the thickness ($z$) direction. This is the case for piezoceramics commonly used for metamaterials \cite{chikh2019piezoelectric,kherraz2018hybridization,kherraz2019tunable} and transducers\cite{wilson1988introduction} (e.g. lead zirconium titanate (PZT) and barium titanate). We assume that the waves in the ($x,y$) plane are decoupled from the waves in the ($x,z$) plane \cite{kherraz2019tunable}, and only include waves in the ($x,z$) plane in our analysis. 

For the electroded cells, which we herein call covered cells, we use a modified 2D version of the approximate model for a covered piezoelectric transducer detailed in \cite{lamberti1990two,lamberti1995general,lamberti2000two}. This model was originally created for standalone transducers and composite piezoelectric/polymer transducers, and we extend it here for plates made entirely of piezoelectric material. We assume the dielectric displacement vector components $D_1=D_2=0$ and $\partial D_3/\partial x= \partial D_3/\partial z=0$. The dielectric displacement in the thickness $z$ direction is assumed to be $D_3=D_0e^{i\omega t}$. The displacement in the $x$ direction is assumed to be
\begin{equation}
\xi_1 = \left[K_1\sin\left(\frac{\omega x}{v_1}\right)+K_2\cos\left(\frac{\omega x}{v_1}\right)\right]e^{i\omega t} ,
\label{eq:xi1}
\end{equation}
while displacement in the $z$ direction is assumed to be
\begin{equation}
\xi_3 = \left[K_3\sin\left(\frac{\omega z}{v_3}\right)+K_4\cos\left(\frac{\omega z}{v_3}\right)\right]e^{i\omega t} ,
\label{eq:xi2}
\end{equation}
where $\omega$ is the angular frequency, and $K_{1-4}$ are constants. When equations (\ref{eq:xi1}) and (\ref{eq:xi2}) are used in the governing equations (\ref{eq:piezo_1})--(\ref{eq:continuity}), along with the relation (\ref{eq:strain}), the phase velocities are found to be
\begin{align}
v_1 = \sqrt{\frac{c^\mathrm{D}_{11}}{\rho}} ; && v_3 = \sqrt{\frac{c^\mathrm{D}_{33}}{\rho}} ,
\end{align}
where
\begin{align}
c^\mathrm{D}_{11} = c^\mathrm{E}_{11} + \frac{e^2_{31}}{\varepsilon^\mathrm{S}_{33}} ; && c^\mathrm{D}_{33} = c^\mathrm{E}_{33} + \frac{e^2_{33}}{\varepsilon^\mathrm{S}_{33}} .
\end{align}
A weak form of the free mechanical boundary condition is used to find constants $K_1$,$K_2$,$K_3$,$K_4$. The lengths $\Sigma_1$, $\Sigma_2$, $\Sigma_3$, and $\Sigma_4$ are along boundaries of a covered cell, as shown in Figure \ref{fig:first_diagram}. The weak form integrals are
\begin{equation}
  \int_{\Sigma_1} T_{11}(x=-a_1/2) \,d\Sigma_1=\int_{\Sigma_2} T_{11}(x=a_1/2) \,d\Sigma_2=0
\end{equation}
\begin{equation}
  \int_{\Sigma_3} T_{33}(z=-h/2) \,d\Sigma_3 = \int_{\Sigma_4} T_{33}(z=h/2) \,d\Sigma_4=0 .
\end{equation}
Due to the lack of bending stress and the small width dimension $a_1$, terms with $e_{15}$ and $e_{24}$ are neglected. In order to find $\partial \phi/\partial x=0$, the (${x,z}$) plane piezoelectric coefficient is set to zero ($e_{31}=0$) \cite{lamberti1995general}. This gives an expression for the electric field in the thickness direction \cite{lamberti1995general}
\begin{equation}
E_3=\frac{D_3}{\varepsilon^\mathrm{S}_{33}}\left\lbrace1-e_{33}\frac{\kappa\omega}{v_1}\left[\cos\left(\frac{\omega z}{v_3}\right) + \tan\left(\frac{\vartheta_3}{2}\right)\sin\left(\frac{\omega z}{v_3}\right)\right]\right\rbrace ,
\label{eq:electric_field}
\end{equation}
where
\begin{equation}
\kappa=\frac{v_3 h c^\mathrm{E}_{11} h_{33} \omega a_1}{c^\mathrm{E}_{11}c^\mathrm{D}_{33}\omega^2 h a_1-4(c^\mathrm{E}_{13})^2 v_1 v_3 \tan(\vartheta_1 / 2)\tan(\vartheta_3 / 2)} .
\end{equation}
The symbols and ratios $\vartheta_1=\omega a_1/v_1$, $\vartheta_3=\omega h/v_3$, and $h_{33}=e_{33}/\varepsilon^\mathrm{S}_{33}$ are kept consistent with \cite{lamberti1995general}. The elastic velocity at the boundaries of the cell are
\begin{align}
  \dot{\xi}_1(x=-a_1) = u_1 ; && \dot{\xi}_1(x=+a_1)=u_2
\end{align}
\begin{align}
  \dot{\xi}_3(z=-h/2) = u_3 ; && \dot{\xi}_3(z=+h/2)=u_4 .
\end{align}
External forces applied to the cell are related to the velocities using the following integration:
\begin{align}
  \int_{\Sigma_1} T_{11}(x=-a_1/2) \,d\Sigma_1=-F_1 && \int_{\Sigma_2} T_{11}(x=a_1/2) \,d\Sigma_2=-F_2
\end{align}
\begin{align}
	\int_{\Sigma_3} T_{33}(z=-h/2) \,d\Sigma_3=-F_3 && \int_{\Sigma_4} T_{33}(z=h/2) \,d\Sigma_4=-F_4 .
\end{align}
This gives equations for the side forces per unit length
\begin{equation}
F_1=b_1u_1+b_2u_2+b_5(u_3+u_4) ,
\label{eq:first_force}
\end{equation}
\begin{equation}
F_2=b_2u_1+b_1u_2+b_5(u_3+u_4) ,
\end{equation}
and top and bottom forces per unit length
\begin{equation}
F_3=b_5(u_1+u_2)+b_3u_3+b_4u_4 ,
\end{equation}
\begin{equation}
F_4=b_5(u_1+u_2)+b_4u_3+b_3u_4 .
\end{equation}
The voltage across the electrodes can be found by integrating (\ref{eq:electric_field}) to give
\begin{equation}
V=b_6(u_3+u_4)+Z_{\mathrm{eq}} I ,
\label{eq:first_voltage}
\end{equation}
where $I$ is the current per unit length, and the electrical impedance of the plate is
\begin{equation}
Z_{\mathrm{eq}} = Z_{\mathrm{C0}} = \frac{1}{i \omega C_0} ,
\label{eq:electrical_impedance1}
\end{equation}
with the clamped capacitance per unit length as
\begin{equation}
C_0 = \frac{\varepsilon_{33}^\mathrm{S} a_1}{h} .
\label{eq:C0}
\end{equation}
The coefficients in the above equations (\ref{eq:first_force})--(\ref{eq:first_voltage}) are
\begin{align}
b_1=\frac{Z_1}{i\tan(\vartheta_1)}; && b_2=\frac{Z_1}{i\sin(\vartheta_1)}
\end{align}
\begin{align}
b_3=\frac{Z_3}{i \tan(\vartheta_3)}; && b_4=\frac{Z_3}{i \sin(\vartheta_3)}
\end{align}
\begin{align}
b_5=\frac{c_{13}^\mathrm{E}}{i \omega}; && b_6=\frac{h_{33}}{i \omega} ,
\end{align}
where the acoustic impedance per length in the $x$ direction is $Z_1=\rho h v_1$ and the acoustic impedance per length in the $z$ direction is $Z_3=\rho a_1 v_3$. These equations differ from the equations formulated in \cite{lamberti1995general} in that they are for 2D plane strain, and use forces per unit length. Equations (\ref{eq:first_force})--(\ref{eq:first_voltage}) can be formulated into a $5\times5$ impedance matrix corresponding to a 5 port element. When a shunted circuit with a complex impedance load $Z_\mathrm{L}$ is connected to the electrodes, The electrical impedance in equation (\ref{eq:electrical_impedance1}) becomes
\begin{equation}
Z_{\mathrm{eq}}=\frac{Z_\mathrm{L} Z_{\mathrm{C0}}}{Z_\mathrm{L}+Z_{\mathrm{C0}}} .
\label{eq:equivalent_impedance}
\end{equation}

In order for covered cells to couple with neighboring uncovered cells on both sides, the voltage port must be duplicated. As shown in the left-hand-side of Figure \ref{fig:mutual_capacitances} (a), the 5 port covered cell element has only one voltage port, which is closed with load $Z^j_\mathrm{L}$. To create a cell with voltage ports on both sides, the one voltage port is duplicated, giving $V^\mathrm{A}_1$ and $V^\mathrm{A}_2$, which are the left side port voltage and the right side port voltage. The current per unit length from the left side is $I^\mathrm{A}_1$, while the current per unit length from the right side is $I^\mathrm{A}_2$. The voltages $V^\mathrm{A}_1$ and $V^\mathrm{A}_2$, along with forces $F_i$, $i=1,2,3,4,5,6$, constitute a 6 port element that can be connected on both the left and right side in a 1D network.

\begin{figure}[!ht]
	\includegraphics[scale=1]{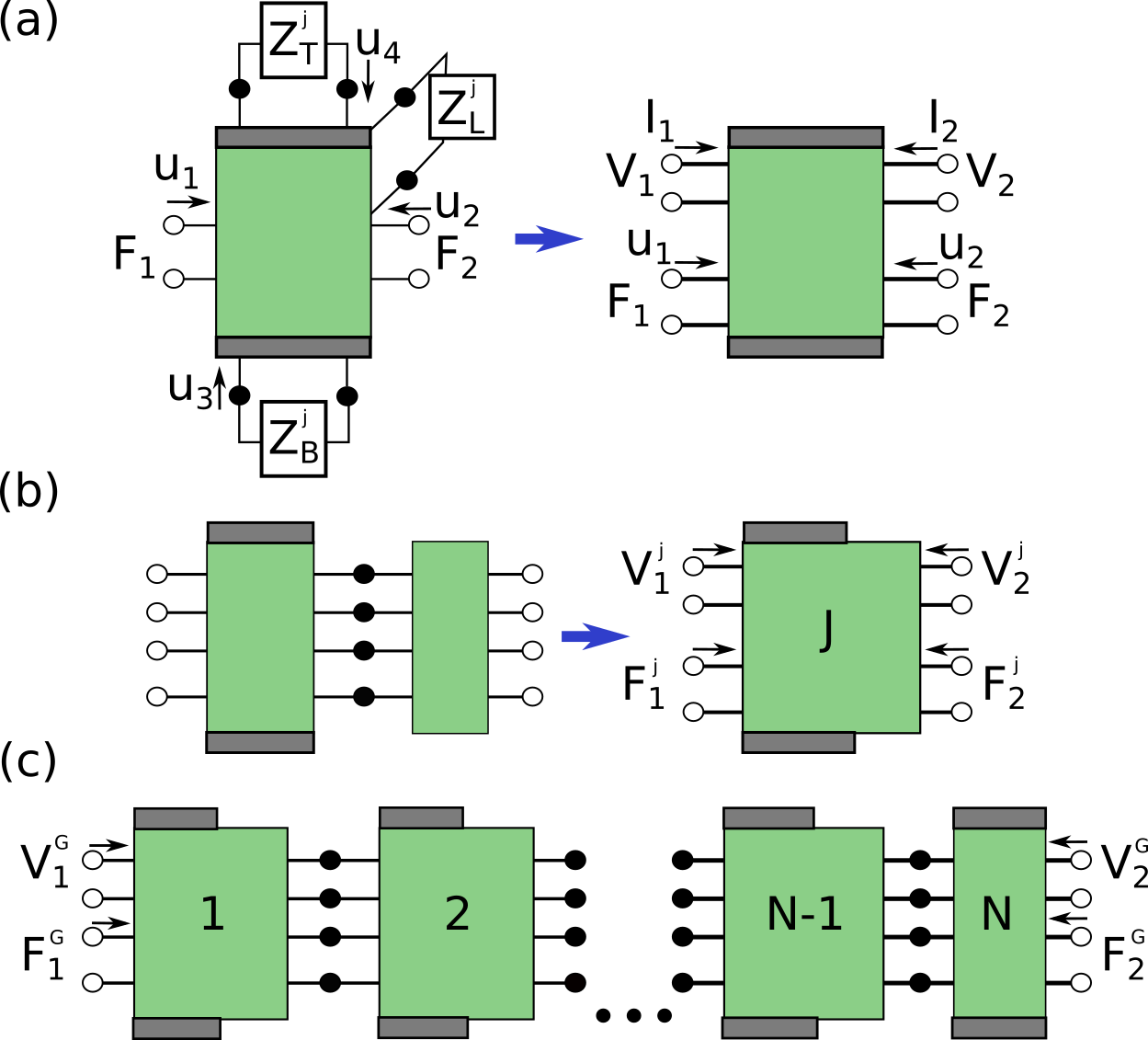} 
	\caption{\label{fig:mutual_capacitances}:  (a) A covered section of the plate shown symbolically as a cell with 5 ports, simplified into a covered cell with 4 ports arranged symmetrically. (b) An uncovered cell of width $a_2$ is connected to a covered cell to form one 4 port unit cell. (c) A 1D network of $N-1$ unit cells, connected to a covered cell to form a global 4 port cell expressing the dynamics of an array of electrode pairs on a piezoelectric plate.}
 \label{fig:mutual_capacitances}
\end{figure}

A 6 port covered cell element can be further simplified to a 4 port element when there is no material on the top or bottom of the plate that requires element modeling. The force at the bottom of the cell can be related to the velocity at the bottom of the cell by $F_3=Z_\mathrm{B} u_3$, where $Z_\mathrm{B}$ is the mechanical impedance of the layer below the plate. Likewise, the force at the top of the cell can be related to the velocity at the top of the cell by $F_4=Z_\mathrm{T} u_4$, where $Z_\mathrm{T}$ is the mechanical impedance of the layer above the plate. This eliminates the need for 2 of the 6 ports, giving the 4 port element shown in Figure \ref{fig:mutual_capacitances}(a). Equations (\ref{eq:first_force})--(\ref{eq:first_voltage}) can be expressed in the impedance matrix form $q=A r$, where $A$ is the impedance matrix, and $r$ is the velocity/current vector, and $q$ is the force/voltage vector. This impedance matrix formulation is fully expressed as
\begin{equation}
\begin{pmatrix}
F_1\\
V_1\\
F_2\\
V_2\\
\end{pmatrix}=
\begin{pmatrix}
b_1-b_5^2b_7 & b_5b_6b_7 & b_2-b_5^2b_7 & b_5b_6b_7 \\
b_5b_6b_7  & Z_{\mathrm{eq}}+b_6^2b_7 & b_5b_6b_7  & Z_{\mathrm{eq}}+b_6^2b_7\\
b_2-b_5^2b_7 & b_5b_6b_7 & b_1-b_5^2b_7 & b_5b_6b_7 \\
b_5b_6b_7  & Z_{\mathrm{eq}}+b_6^2b_7 & b_5b_6b_7  & Z_{\mathrm{eq}}+b_6^2b_7\\
\end{pmatrix}
\begin{pmatrix}
u_1\\
I_1\\
u_2\\
I_2\\
\end{pmatrix}
\end{equation}
with the use of an additional coefficient
\begin{equation}
b_7=\frac{2(b_3-b_4)+Z_\mathrm{B}-Z_\mathrm{T}}{b_3^2-b_4^2+b_4Z_\mathrm{B}-b_3Z_\mathrm{T}} .
\end{equation}

To formulate an impedance matrix equation for open (uncovered by electrodes) cells, electrostatic lumped impedances are implemented. Within open cells, there is a non-negligible component of the electric field in the $x$ direction \cite{kherraz2018hybridization}. We therefore assume a non-zero dielectric displacement in the $x$ direction $D_1$, and $\partial D_1/\partial x=\partial D_3/\partial z=0$ to satisfy (\ref{eq:dielectric_displacement}). However, even if the effects of $D_3$ are neglected, equations like (\ref{eq:first_force})--(\ref{eq:first_voltage}) cannot be formulated with the same assumed solutions for the displacements. Therefore, we simplify the problem by neglecting electro-mechanical coupling. This gives a linear impedance matrix
\begin{equation}
\mathbf{B}=
\begin{pmatrix}
b_1-b_5^2b_7 & 0 & b_2-b_5^2b_7 & 0\\
0 & Z^\mathrm{E}_{11} & 0 & Z^\mathrm{E}_{12}\\
b_2-b_5^2b_7 & 0 & b_1-b_5^2b_7 & 0\\
0 & Z^\mathrm{E}_{21} & 0 & Z^\mathrm{E}_{22}\\
\end{pmatrix} .
\label{eq:uncovered_cell_impedance}
\end{equation}
In the above matrix, null values $B_{12}=B_{14}=B_{21}=B_{23}=B_{32}=B_{34}=B_{41}=B_{43}=0$ are the electro-mechanical coupling terms. The potential difference between the top left corner and bottom left corner of the cell is $V^\mathrm{B}_1=Z^\mathrm{E}_{11}I^\mathrm{B}_1+Z^\mathrm{E}_{12}I^\mathrm{B}_2$, while the potential difference between the top right corner and bottom right corner of the cell is $V^\mathrm{B}_2=Z^\mathrm{E}_{21}I^\mathrm{B}_1+Z^\mathrm{E}_{22}I^\mathrm{B}_2$. The current per unit length going into the cell from the left is $I^\mathrm{B}_1$, while the current per unit length going into the cell from the right is $I^\mathrm{B}_2$. We calculate the electrical impedances $Z^\mathrm{E}_{11}$, $Z^\mathrm{E}_{12}$, $Z^\mathrm{E}_{21}$, and $Z^\mathrm{E}_{22}$ by estimating mutual capacitance per unit length. 

The placement of mutual capacitance captures the first two modes of electro-mechanical propagation. The first mode of propagation has electric fields between parallel electrodes opposite in polarity with neighboring electrode pairs, as shown in Figure \ref{fig:cap_diag} (a). This occurs when $\lambda=\Lambda$ \cite{kherraz2018hybridization}. The capacitance between nearest neighbor top electrodes is labeled in Figure \ref{fig:cap_diag} (a) as $C_1$. To estimate $C_1$, we use an analytical method based on conformal mapping \cite{stellari2000new} which gives 
\begin{equation}
C_1 = \frac{\varepsilon_{\mathrm{ave}}}{4}\mathcal{M}(k_\mathrm{T}(\mu))
\label{eq:C1}
\end{equation}
where $\varepsilon_{\mathrm{ave}}=(\varepsilon_{11}+\varepsilon_{33})/2$, and $\mu=a_1/a_2$. The function $\mathcal{M}$ is
\begin{equation}
\mathcal{M}\left(k(z,y)\right)\cong
	\begin{cases} 
      \frac{2\pi}{\ln\left[2\frac{1+\left(1-k_{\mathrm{BCP}}^2\right)^{1/4}}{1-\left(1-k_{\mathrm{BCP}}^2\right)^{1/4}}\right]} & 0\leq k_{\mathrm{BCP}} \leq \frac{1}{\sqrt{2}} \\
      \frac{2}{\pi}\ln\left[2\frac{1+\sqrt{k_{\mathrm{BCP}}}}{1-\sqrt{k_{\mathrm{BCP}}}}\right] & \frac{1}{\sqrt{2}} \leq k_{\mathrm{BCP}} \leq 1 
  \end{cases},
\end{equation}
and $k_{\mathrm{S}}$ is a dimensionless geometric parameter calculated by
\begin{equation}
k_{\mathrm{S}}(z,y) = \frac{\sqrt{\sinh\left[\frac{\pi}{2}(z+y)\right]\sinh\left(\frac{\pi}{2}z\right)}}{\cosh\left[\frac{\pi}{2}\left(z+\frac{y}{2}\right)\right]}.
\end{equation}
The capacitance between the top electrodes and the aligned bottom electrodes is partially taken into consideration by Eq. (\ref{eq:C0}). However, this equation underestimates the capacitance between parallel plates in a continuous medium by neglecting fringing effects. To more accurately estimate parallel plate capacitance we use \citep{majumdar2010alignment}

\begin{equation}
C_\mathrm{p} = \frac{1.15\varepsilon_{\mathrm{ave}}a_1}{h}+1.4\varepsilon_{\mathrm{ave}}(a_1+1)\left( \frac{2T}{h}\right)^{0.222}+1.03\varepsilon_{\mathrm{ave}}h\left(\frac{2T}{h}\right)^{0.728}
\label{eq:Cp}
\end{equation}
In the above equation, the first term is the parallel plate capacitance with a multiplicative factor of $1.15$, while the second and third terms account for fringe effects on both sides of the electrodes. The fringe capacitance $C_2$ within an open cell is taken as $C_2=(C_\mathrm{p}-C_0)/2$. 

The second mode of electromechanical propagation is when neighboring pairs of electrodes have electric fields with the same polarity, as shown in Figure \ref{fig:cap_diag} (b). This occurs when $\lambda\gg\Lambda$ \cite{kherraz2018hybridization}. In this mode, $C_1$ can be neglected. However, there is a non-negligable capacitance between top electrodes and the nearest neighbor bottom electrodes, labeled as $C_3$ in Figure \ref{fig:cap_diag} (b). We use a modified version of the parallel plate capacitance to account for the misalignment of the electrodes, which is

\begin{equation}
C_3 = \frac{\varepsilon_{\mathrm{ave}}a_1}{d}+1.4\varepsilon_{\mathrm{ave}}(a_1+1)\left( \frac{2T}{d}\right)^{0.222}+1.03\varepsilon_{\mathrm{ave}}h\left(\frac{2T}{d}\right)^{0.728}
\label{eq:C3}
\end{equation}
where $d=\sqrt{h^2+(a_2+a_1)^2}$ is the adjusted distance between the electrodes. 

Lumped impedances with both $C_1$ and $C_3$ accounts for both of these modes. The result is a lattice circuit, between two shunted impedance circuits, as shown in Figure \ref{fig:cap_diag} (c). The electrical impedance parameters are
\begin{equation}
Z^\mathrm{E}_{11}=\frac{Z_3Z_2^2(Z_1+2Z_3)(Z1+Z_3)+Z_2Z_3^2(Z_1^2+2Z_1Z_3)}{2Z_2Z_3(Z_1+Z_3)(Z_1+2Z_3)+Z_2^2(Z_1+2Z_3)^2+Z_3^2(Z_1^2+2Z_1Z_3)};
\label{eq:ZE11}
\end{equation}
\begin{equation}
Z^\mathrm{E}_{12}=\frac{Z_2^2(Z_1+2Z_3)^2+Z_3^2(Z_1+2Z_3)}{2Z_2Z_3(Z_1+2Z_3)(Z_1+Z_3)+Z_2^2(Z_1+2Z_3)^2+Z_3^2(Z_1^2+2Z_1Z_3)};
\label{eq:ZE12}
\end{equation}
where $Z_1=1/(i\omega C_1)$, $Z_2=1/(i\omega C_2)$, and $Z_3=1/(i\omega C_3)$. The other terms in the symmetric matrix are $Z^\mathrm{E}_{22}=Z^\mathrm{E}_{11}$ and $Z^\mathrm{E}_{21}=Z^\mathrm{E}_{12}$.

 \begin{figure}[!ht]
 \includegraphics[scale=1]{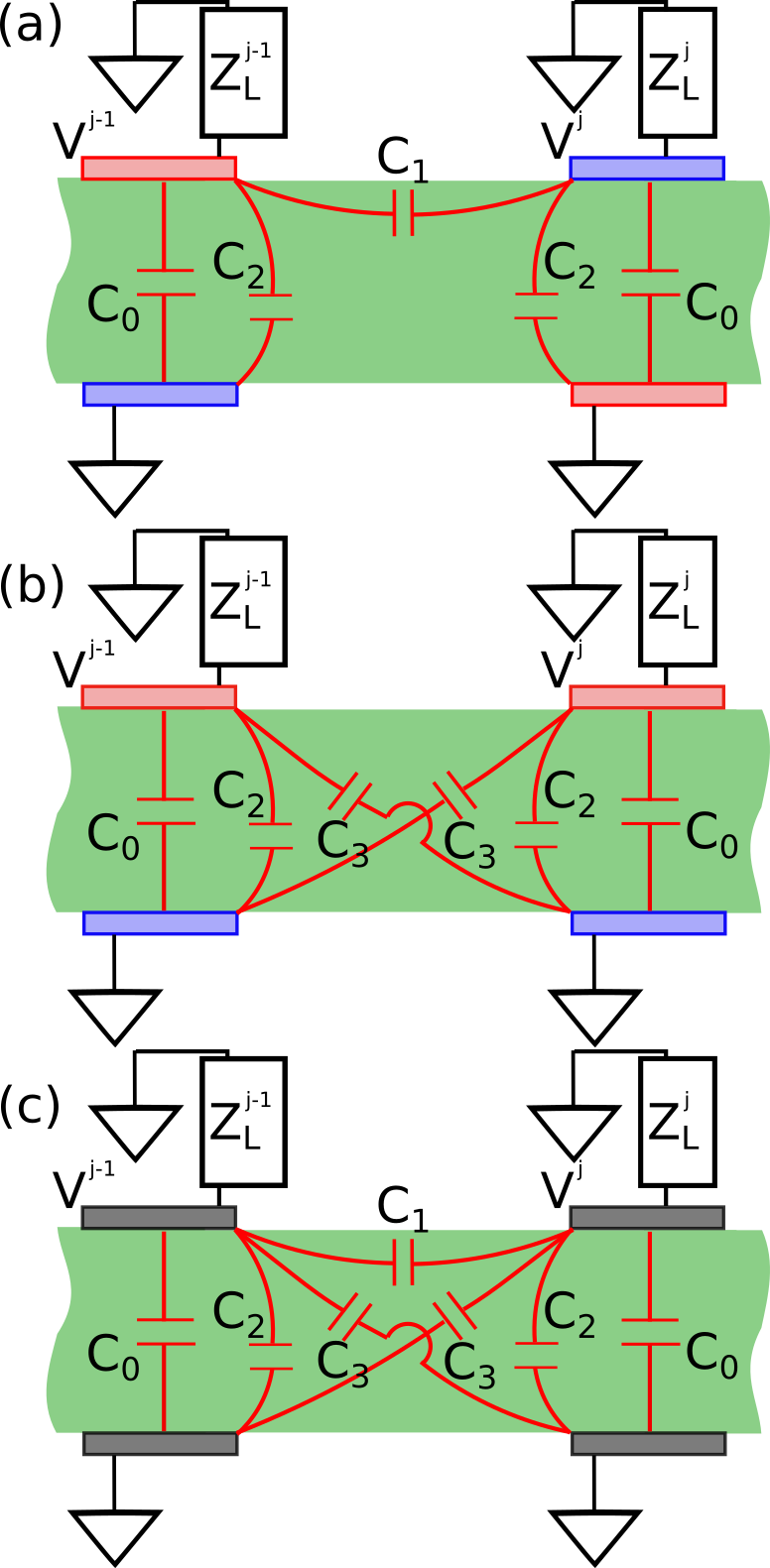} 
 \label{fig:cap_diag}
 \caption{\label{fig:cap_diag}: (a) Cross-sectional diagrams of a segment of the piezoelectric plate with two electrode pairs and mutual capacitances between electrodes, all shown in red. (a) The first electrical mode with capacitance between top and bottom electrodes within a cell ($C_0$ and $C_2$), and nearest-neighbor co-planar electrodes ($C_1$). (b) The second electrical mode with capacitances $C_0$ and $C_2$, as wells as top and nearest-neighbor bottom electrodes $C_3$. (c) Capacitances $C_0$,$C_1$,$C_2$, and $C_3$ all taken into account to formulate the impedance matrix Eq. (\ref{eq:ZE11}--\ref{eq:ZE12}).}
\end{figure}

\subsection{Transfer Matrix Method using Scattering Matrices}
To calculate scattering parameters for plates consisting of multiple electrode strips, scattering matrices are used. Scattering matrices are commonly used for the transfer matrix method when modeling electromagnetic wave propagation, due to being numerically stable and intrinsically containing the refection and transmission coefficients \cite{rumpf2011improved}. One can transform the impedance matrix of a covered cell $\mathbf{A}$ into a scattering matrix $\mathbf{S}^\mathrm{A}$ with the formula \cite{kurokawa1965power}
\begin{equation}
\mathbf{S}^\mathrm{A} = \mathbf{G}_{\mathrm{ref}}(\mathbf{A}-\mathbf{Z}^*_{\mathrm{ref}})(\mathbf{A}+\mathbf{Z}_{\mathrm{ref}})^{-1}\mathbf{G}^{-1}_{\mathrm{ref}} .
\end{equation}
The reference impedance matrices are $Z_{\mathrm{ref},ij}=Z_{0,ij}\delta_{ij}$, and $G_{\mathrm{ref},ij}=1/\sqrt{|Z_{0,ij}|}\delta_{ij}$, where $\delta_{ij}$ is the Kronecker delta function. The mechanical reference impedances are assumed to be $Z_{0,11}=Z_{0,33}=\rho h v_1$, while the electrical reference impedances are assumed to be $Z_{0,22}=Z_{0,44}=50$ $\Omega$. Likewise, one can transform the impedance matrix of an uncovered cell $\mathbf{B}$ into a scattering matrix $\mathbf{S}^\mathrm{B}$. 

The S matrix for a symmetric unit cell consists of one uncovered cell and one covered cell, shown in Figure \ref{fig:mutual_capacitances} (c). The scattering matrix of a unit cell can be found with
\begin{equation}
\mathbf{S}^{\mathrm{UC}} = \mathbf{S}^{\mathrm{A}} \otimes \mathbf{S}^{\mathrm{B}} ,
\end{equation}
where $\otimes$ is the Redheffer star product  \cite{rumpf2011improved}, which is detailed in Appendix A.
The scattering matrix $\mathbf{S}^{\mathrm{UC},j}$ for the $j$th unit cell gives the relation
\begin{equation}
\begin{pmatrix}
F^{j,-}_1\\
V^{j,-}_1\\
F^{j,+}_2\\
V^{j,+}_2\\
\end{pmatrix}=
\begin{pmatrix}
S^{j}_{11} & S^{j}_{12} & S^{j}_{13} & S^{j}_{14}\\
S^{j}_{21} & S^{j}_{22} & S^{j}_{23} & S^{j}_{24}\\
S^{j}_{31} & S^{j}_{32} & S^{j}_{33} & S^{j}_{34}\\
S^{j}_{41} & S^{j}_{42} & S^{j}_{43} & S^{j}_{44}\\
\end{pmatrix}
\begin{pmatrix}
F^{j,+}_1\\
V^{j,+}_1\\
F^{j,-}_2\\
V^{j,-}_2\\
\end{pmatrix}
\end{equation}
where superscript $+$ and $-$ denote frontward and backward waves, respectively.

As shown in Figure \ref{fig:mutual_capacitances} (c), a scattering matrix for a section of plate with $N$ electrode pairs can be found by coupling $N-1$ unit cells with a covered cell. This global scattering matrix can be found with
\begin{equation}
\mathbf{S}^{\mathrm{G}} = \mathbf{S}^{\mathrm{UC},1}\otimes \mathbf{S}^{\mathrm{UC},2}\otimes \dots\mathbf{S}^{\mathrm{UC},N-1} \otimes \mathbf{S}^{\mathrm{A}}.
\label{eq:global_scattering}
\end{equation}
If the plate has reflection and transmission regions without electrodes, asymmetric scattering matrices $\mathbf{S}^{\mathrm{RFL}}$ and $\mathbf{S}^{\mathrm{TRS}}$ coupling the cells to the reflection and transmission regions can be used \cite{rumpf2011improved}. Cells in this 2D plate model can each have different electrical and mechanical boundary conditions, as well as different material properties and thickness.

The dispersion relations for infinitely periodic plates can also be evaluated with the scattering matrix of one unit cell $j$. Terms in the scattering matrix $\mathbf{S}^{\mathrm{UC}}$ can be moved around to form a relation between wave amplitudes from unit-cell ports $F^{j,+}_2$, $V^{j,+}_2$ and ports $F^{j,+}_1$, $V^{j,+}_1$, given as
\begin{equation}
\begin{pmatrix}
0 & 0 & -S^{\mathrm{UC}}_{13} & -S^{\mathrm{UC}}_{14}\\
0 & 0 & -S^{\mathrm{UC}}_{23} & -S^{\mathrm{UC}}_{24}\\
1 & 0 & -S^{\mathrm{UC}}_{33} & -S^{\mathrm{UC}}_{34}\\
0 & 1 & -S^{\mathrm{UC}}_{43} & -S^{\mathrm{UC}}_{44}\\
\end{pmatrix}
\begin{pmatrix}
F^{j,+}_2\\
V^{j,+}_2\\
F^{j,-}_2\\
V^{j,-}_2\\
\end{pmatrix}
=
\begin{pmatrix}
S^{\mathrm{UC}}_{11} & S^{\mathrm{UC}}_{12} & -1 & 0\\
S^{\mathrm{UC}}_{21} & S^{\mathrm{UC}}_{22} & 0 & -1\\
S^{\mathrm{UC}}_{31} & S^{\mathrm{UC}}_{32} & 0 & 0\\
S^{\mathrm{UC}}_{41} & S^{\mathrm{UC}}_{42} & 0 & 0\\
\end{pmatrix}
\begin{pmatrix}
F^{j,+}_1\\
V^{j,+}_1\\
F^{j,-}_1\\
V^{j,-}_1\\
\end{pmatrix} .
\end{equation}
The Bloch wave theory for infinitely periodic plates then gives the relation \cite{brillouin1953wave}
\begin{equation}
\begin{pmatrix}
F^{j,+}_2\\
V^{j,+}_2\\
F^{j,-}_2\\
V^{j,-}_2\\
\end{pmatrix}=
e^{-i \beta \Lambda}
\begin{pmatrix}
F^{j,+}_1\\
V^{j,+}_1\\
F^{j,-}_1\\
V^{j,-}_1\\
\end{pmatrix}
\end{equation}
where $\beta$ is the effective wavenumber at a given frequency. This relation allows us to formulate
\begin{equation}
\begin{pmatrix}
0 & 0 & -S^{\mathrm{UC}}_{13} & -S^{\mathrm{UC}}_{14}\\
0 & 0 & -S^{\mathrm{UC}}_{23} & -S^{\mathrm{UC}}_{24}\\
1 & 0 & -S^{\mathrm{UC}}_{33} & -S^{\mathrm{UC}}_{34}\\
0 & 1 & -S^{\mathrm{UC}}_{43} & -S^{\mathrm{UC}}_{44}\\
\end{pmatrix}
\begin{pmatrix}
F^{j,+}_1\\
V^{j,+}_1\\
F^{j,-}_1\\
V^{j,-}_1\\
\end{pmatrix}
= e^{-i \beta \Lambda}
\begin{pmatrix}
S^{\mathrm{UC}}_{11} & S^{\mathrm{UC}}_{12} & -1 & 0\\
S^{\mathrm{UC}}_{21} & S^{\mathrm{UC}}_{22} & 0 & -1\\
S^{\mathrm{UC}}_{31} & S^{\mathrm{UC}}_{32} & 0 & 0\\
S^{\mathrm{UC}}_{41} & S^{\mathrm{UC}}_{42} & 0 & 0\\
\end{pmatrix}
\begin{pmatrix}
F^{j,+}_1\\
V^{j,+}_1\\
F^{j,-}_1\\
V^{j,-}_1\\
\end{pmatrix}
\label{eq:gen_eig_value_prblm}
\end{equation}
which is in the generalized eigenvalue form. This eigenvalue problem can be evaluated to find $\lambda=e^{-i\beta \Lambda}$ for a given frequency. The effective wavenumber $\beta$ can then be found with
\begin{equation}
\cos(\beta \Lambda)=\frac{1}{2}\left(e^{-i\beta\Lambda}+e^{i\beta\Lambda}\right)
\label{eq:effective_wavenumber}
\end{equation}

\subsection{Effective Parameter Retrieval}
Given the scattering parameters for a unit-cell or for an entire plate consisting of multiple cells, the effective elastic impedance and effective elastic wavenumber can be calculated. We use a method of extracting the effective mechanical impedance that is commonly used for electromagnetic metamaterials \cite{arslanagic2013review}. The effective mechanical impedance can be found from the mechanical scattering coefficients with \cite{arslanagic2013review}
\begin{equation}
Z_{\mathrm{eff}}=\pm Z_0 \sqrt{\frac{(S_{11}+1)^2-S_{31}^2}{(S_{11}-1)^2-S_{31}^2}}
\label{eq:equivalent_impedance}
\end{equation}
where $Z_0$ is the reference impedance, which we set as equal to the mechanical impedance per length $Z_0=Z_{0,11}$.

The effective wavenumber $k_{\mathrm{eff}}$ is found by the definition $P_\mathrm{A}=e^{ik_{\mathrm{eff}}N\Lambda}$. The complex value $P_\mathrm{A}$ can be found at a given frequency by \cite{arslanagic2013review}
\begin{equation}
P_\mathrm{A}=\frac{S_{31}(Z_{\mathrm{eff}}+Z_0)}{(Z_{\mathrm{eff}}+Z_0)-S_{11}(Z_{\mathrm{eff}}-Z_0)}
\end{equation}
The effective wavenumber is then calculated by \cite{arslanagic2013review}
\begin{equation}
k_{\mathrm{eff}}=\frac{1}{N\Lambda}\left[-\left(\arg(P_\mathrm{A})+2\pi q\right)+i\log|P_\mathrm{A}|\right];
\label{eq:equivalent_wavenumber}
\end{equation}
where $q$ is the branch index that must be chosen for each frequency.
\section{Numerical Results}
\label{sec:numerical}
Band diagrams and mutual capacitances, calculated using the analytical methods in Section \ref{sec:methods}, were compared with calculations from finite element analysis (FEA) performed in COMSOL multiphysics \citep{COMSOL:2019}. A shunted circuit load of pure inductance values $L=0.1$ mH and $L=1$ mH were used for the calculations presented. For all of the numerical studies in this publication, the properties of PZT4, given in Table \ref{table:piezo_values}, were used. PZT4 was chosen due to its prevalence in transducer applications \cite{wilson1988introduction}. The COMSOL \cite{COMSOL:2019} tabulated property values for PZT4 were used, which are only slightly different from the often cited measurements by Vernitron, Inc \cite{wilson1988introduction}. The low-frequency hybridization band gap that appears in the presented results is easily tunable with an adjustment to the impedance $Z_\mathrm{L}$. Further, the hybridization band gap is below the homogenization limit leading to a metamaterial, compared with the higher frequency Bloch wave band gap. Therefore, we focus our comparison of the FEA and analytical results on the hybridization band gap.

\begin{table}[h!]
\caption{\label{table:piezo_values}: Property values for PZT4 \cite{COMSOL:2019} used for all the results presented in this publication. The permittivity of free space is $\varepsilon_0=8.85418782\times10^{-12}$.}
\centering
\begin{tabular}{ |c|c|c| } 
\hline
 Property & Value & Unit \\ [0.5ex] 
 \hline\hline
$\rho$ & 7700 & $\mathrm{kg/m^{3}}$\\
$c_{11}^\mathrm{E}$ & 175 & GPa \\
$c_{13}^\mathrm{E}$ & 95.0 & GPa\\
$c_{33}^\mathrm{E}$ & 124 & GPa\\
$e_{31}$ & -2.62 & $\mathrm{C/m^{2}}$\\
$e_{33}$ & 16.48 & $\mathrm{C/m^{2}}$\\
$e_{15}$ & 10.00 & $\mathrm{C/m^{2}}$\\
$e_{24}$ & 10.00 & $\mathrm{C/m^{2}}$\\
$\varepsilon_{11}^\mathrm{S}$ & 800.0$\varepsilon_0$ & $\mathrm{s^4 A^2/(m^{3}kg^{1})}$\\
$\varepsilon_{33}^\mathrm{S}$ & 767.6$\varepsilon_0$ & $\mathrm{s^4 A^2/(m^{3}kg^{1})}$\\
\hline
\end{tabular}
\label{table:piezo_values}
\end{table}

\subsection{Finite Element Analysis}
Eigenfrequency FEA studies were conducted with and without altered property values to estimate the effect of some of the key assumptions made in the analytical model formulation. Studies with and without certain piezoelectric coefficients set to zero ($e_{15}=e_{31}=0$) were conducted. A 2D rectangular piezoelectric solid with perfectly conductive line electrodes on the top and bottom was used.  A triangular mesh with a maximum dimension $\delta=\lambda_\mathrm{m}/200$, where $\lambda_\mathrm{m}$ is the smallest wavelength in the study, was utilized to divide the occupied geometry. Shunted electrical circuits with passive elements were accounted for by lumped parameter differential equations coupled to the partial differential equations for the piezoelectric solid. The Bloch wave condition was used for the right and left boundaries of a unit-cell, allowing the calculation of eigenfrequencies for an infinitely periodic plate. A pair of band diagrams for $h=2$ mm, $a_1=1$ mm, $a_2=2$ mm, and $L=0.1$ mH, with and without $e_{15}=e_{31}=0$ is shown in Figure \ref{fig:band_diag_compare}.

  \begin{figure}[!ht]
  \includegraphics[scale=1]{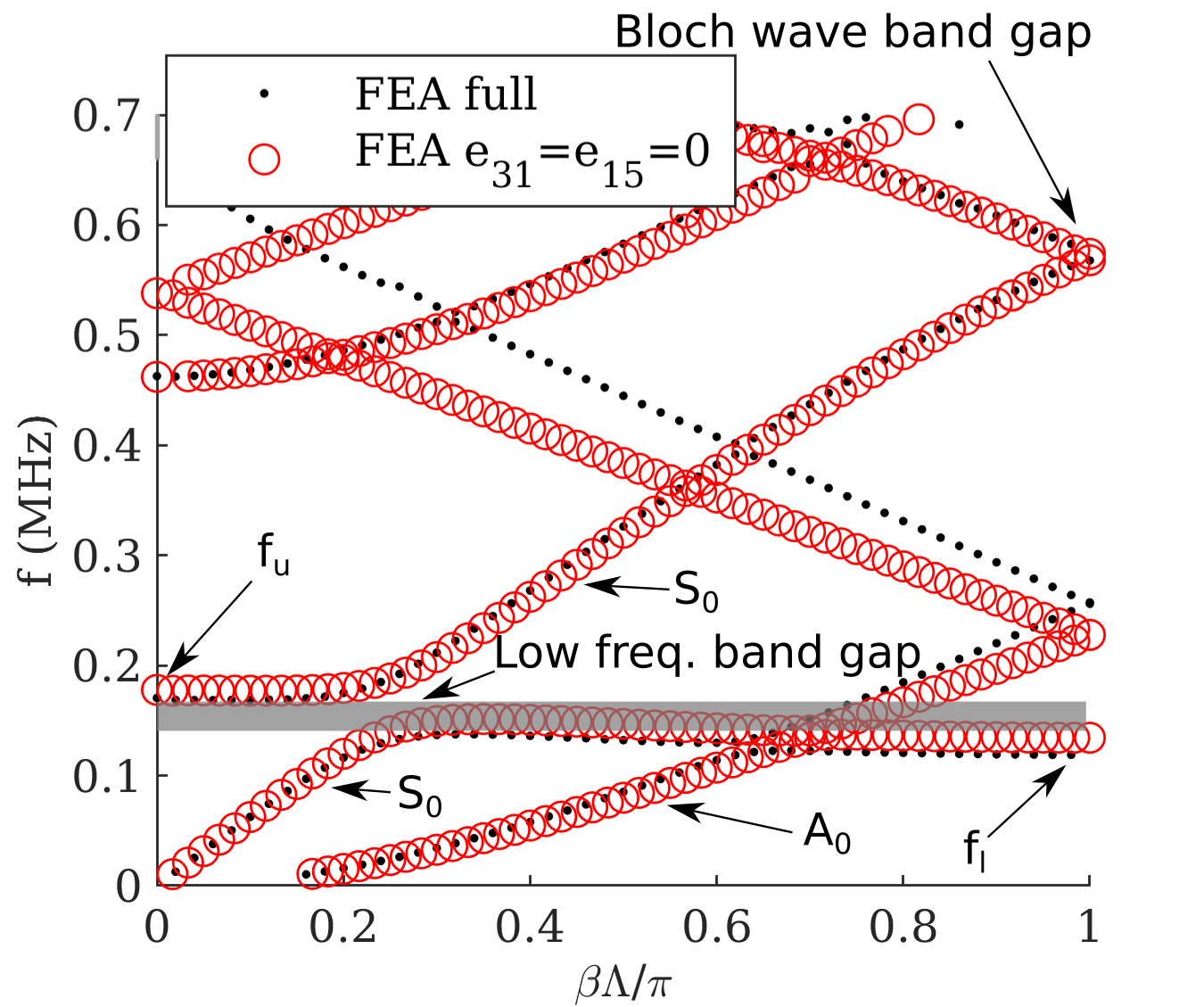} 
 \label{fig:band_diag_compare}
 \caption{\label{fig:band_diag_compare}: Band diagrams constructed from eigenfrequency solutions using FEA. The solutions given by red circles neglect selected terms in the governing equations by setting $e_{31}=e_{15}=0$, while the solutions in black come from calculations without neglecting terms. The low frequency band gap is shown with a gray rectangle. The zeroth order symmetric ($S_0$) and asymmetric $A_0$ modes are labeled. The $S_0$ mode at $k=0$, which sometimes coincides with the upper frequency of the hybridization band gap, is labeled as $f_\mathrm{u}$. The $S_0$ mode at $k=1$, which in this example does not coincide with the lower end of the low frequency band gap, is labeld as $f_\mathrm{l}$.  }
 \end{figure}
 
As shown in Figure \ref{fig:band_diag_compare}, the largest impact in the dispersion relation made by setting $e_{31}=e_{15}=0$ is found in the zeroth order asymmetric lamb waves $A_0$. This is due to the absence of the $\epsilon_{31}$ and $\epsilon_{15}$ strain terms. The analysis with the neglected terms estimates a higher low-frequency hybridization band gap for $S_0$, with a gap that spans 151--177 kHz, compared to 137--169 kHz without any terms neglected. Neglecting piezoelectric coefficients ($e_{31}=e_{15}=0$) produces a slight change in the hybridization band gap, but does not change the overall shape and structure of the symmetric mode $S_0$ band diagram.

\subsection{Band Diagrams using the Analytical Model}
Methods in Section \ref{sec:methods} were used to numerically calculate band diagrams for infinitely periodic plates. It was assumed that the plates were mechanically free on the top and bottom ($Z_\mathrm{T}=0$, $Z_\mathrm{B}=0$). These band diagrams were compared with band diagrams calculated using full FEA with $e_{15}\neq0$ and $e_{31}\neq0$. Line electrode elements were used in the mesh for the FEA calculations. A periodic condition was used for both the mechanical and electrostatic boundary on the left and right-hand side of the geometry. Equations (\ref{eq:gen_eig_value_prblm}) and (\ref{eq:effective_wavenumber}) were solved numerically, giving effective elastic wavenumbers $\beta$ at frequencies $f\in {0.01,0.7}$ MHz, where $f=\omega/(2\pi)$. These solutions are shown with red $+$ markers in Figure \ref{fig:band_diag_2D} for $h=2$ mm and $a_1=1$ mm, $a_2=2$ mm, and L=$0.1$ mH, while the FEA solutions are shown with black dots.  

\begin{figure}[!ht]
	\includegraphics[scale=1]{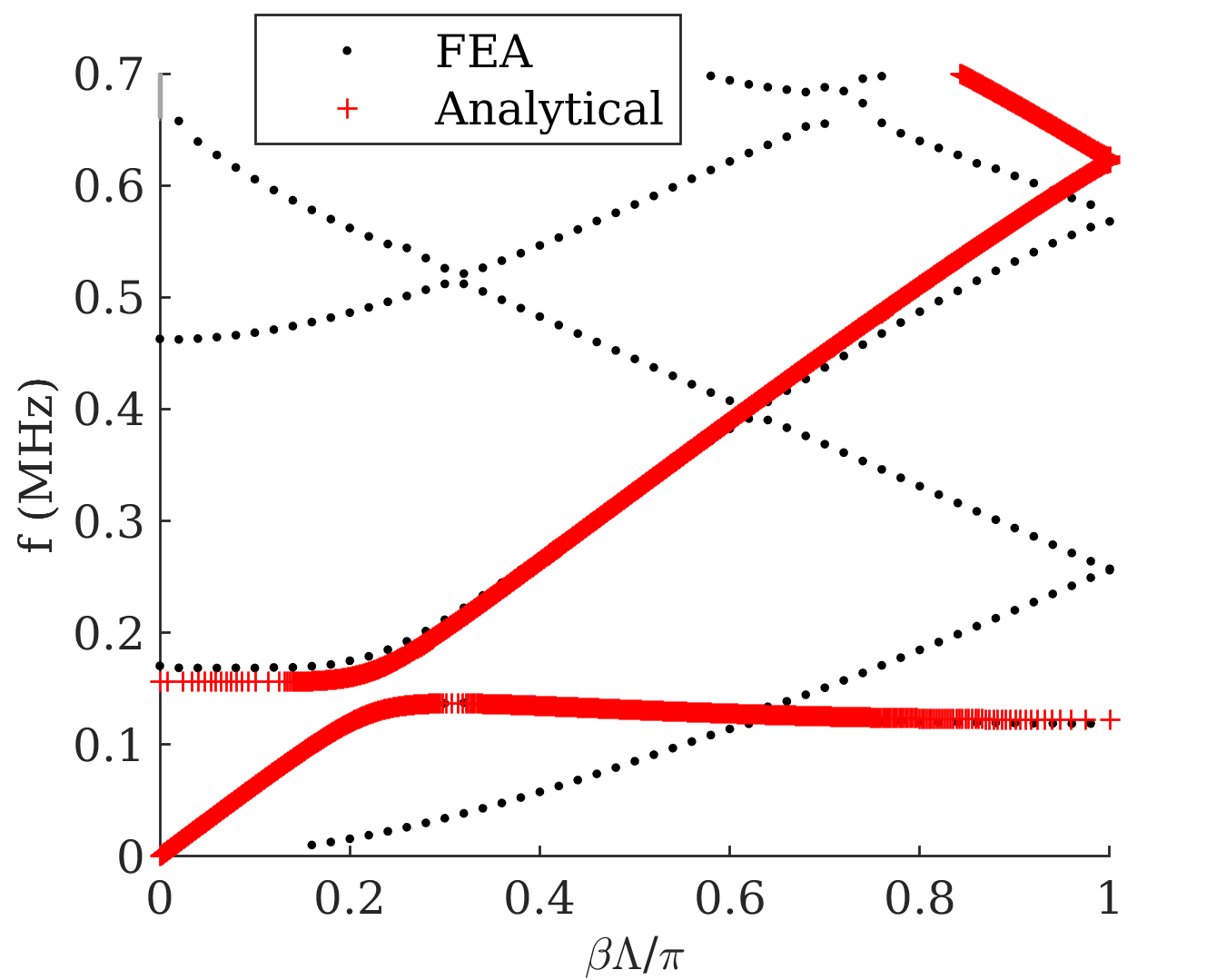} 
 \label{fig:band_diag_2D}
 \caption{\label{fig:band_diag_2D}: Band diagram using both FEA and the approximate analytical method showing the real part 	of the effective elastic wavenumber $\beta$ vs. eigenfrequency $f$.}
\end{figure}

As shown in Figure \ref{fig:band_diag_2D}, the analytical results compare better with the FEA results near the hybridization band gap than the high frequency Bloch wave band gap. The analytical method estimates the Bloch wave band gap to be 620--625 kHz, while the FEA results estimate it to be 568--574 kHz. We define the percent difference in the middle frequency of the hybridization bandgap as
\begin{equation}
\Delta f_\mathrm{m}=100\%\times\frac{f^\mathrm{A}_\mathrm{m}-f^\mathrm{F}_\mathrm{m}}{f^\mathrm{F}_\mathrm{m}} ,
\label{eq:percent_difference}
\end{equation}
where subscripts $A$ and $F$ refer to the analytical method and the FEA results, respectively. For this example, $\Delta f_\mathrm{m}=-6.7$kHz (-4.4\%). The percent difference in the width of the hybridization band gap $\Delta f_\mathrm{g}$, which is calculated with Eq. (\ref{eq:percent_difference}) using $f^\mathrm{A}_\mathrm{g}$ and $f^\mathrm{F}_\mathrm{g}$, in this example is 11.9 kHz (-37\%). The eigenfrequencies at $k=1$ ($f_\mathrm{l}$) and $k=0$ ($f_\mathrm{u}$) also show some disagreement. The percent differences $\Delta f_\mathrm{u}$ and $\Delta f_\mathrm{l}$ were calculated using  Eq. (\ref{eq:percent_difference}) with $f^\mathrm{A}_\mathrm{u}$,$f^\mathrm{F}_\mathrm{u}$,$f^\mathrm{A}_\mathrm{l}$, and $f^\mathrm{F}_\mathrm{l}$. For the example geometry used for Figure \ref{fig:band_diag_2D}, $\Delta f_\mathrm{l}=2.7\%$ and $\Delta f_\mathrm{u}=-7.2\%$. The analytical method underestimates the band gap by placing it at lower frequencies than those found from FEA, and underestimates the width of the band gap, but does very well capturing the electromechanical coupling, the overall band structure, and the hybridization band gap.

The example band diagrams in Figure \ref{fig:band_diag_compare} and Figure \ref{fig:band_diag_2D}, show that while the disagreement between the analytical method and FEA is partially due to the neglected piezoelectric coefficients $e_{31}=e_{15}=0$, it is also due to other factors. The FEA band diagram in \ref{fig:band_diag_compare} is at slightly higher freqencies when $e_{31}=e_{15}=0$. However, the band diagram calculated using the approximate analytical method is at slightly lower frequencies than the band diagram from FEA. The decoupling of the electrical and mechanical impedances in Equation (\ref{eq:uncovered_cell_impedance}), and the use of an approximate electrical capacitance for electrical impedance (Eq. (\ref{eq:C1})--(\ref{eq:C3}) also seem to cause discrepancies between band diagrams calculated using FEA and the analytical method.

In order to understand the efficacy of the analytical model in predicting what frequencies the hybridization band gap occurs, the percent difference $\Delta f_\mathrm{m}$ was calculated for different plate thickness values $h$ and different gap values between the electrodes $a_2$. In Figure \ref{fig:band_diag_parameters} (a), the $\Delta f_\mathrm{m}$ values are shown as percentages for $a_2 \in \{0.8,...,3.2\}$mm and $h \in \{2,3,4\}$mm. In Figure \ref{fig:band_diag_parameters} (b)-(d), 
It can be seen from Figure \ref{fig:band_diag_parameters} (a) that $\Delta f_\mathrm{m}$ is greater for larger $h$ values, and less for lower $a_2$ values. The $|\Delta f_\mathrm{m}|$ values are all below $13\%$ for these geometries, and are as low as $0.45\%$, demonstrating a close agreement between the approximate analytical method and multiphysics FEA for these geometries. The fact that $\Delta f_\mathrm{m}$ is not consistently positive likely has to due with the accuracy of the capacitance estimations and/or the use of Eq. (\ref{eq:ZE11} and (\ref{eq:ZE12}) for estimating $Z^\mathrm{E}_{11}$ and $Z^\mathrm{E}_{12}$.

\begin{figure}[!ht]
 \centering
	\includegraphics[scale=1]{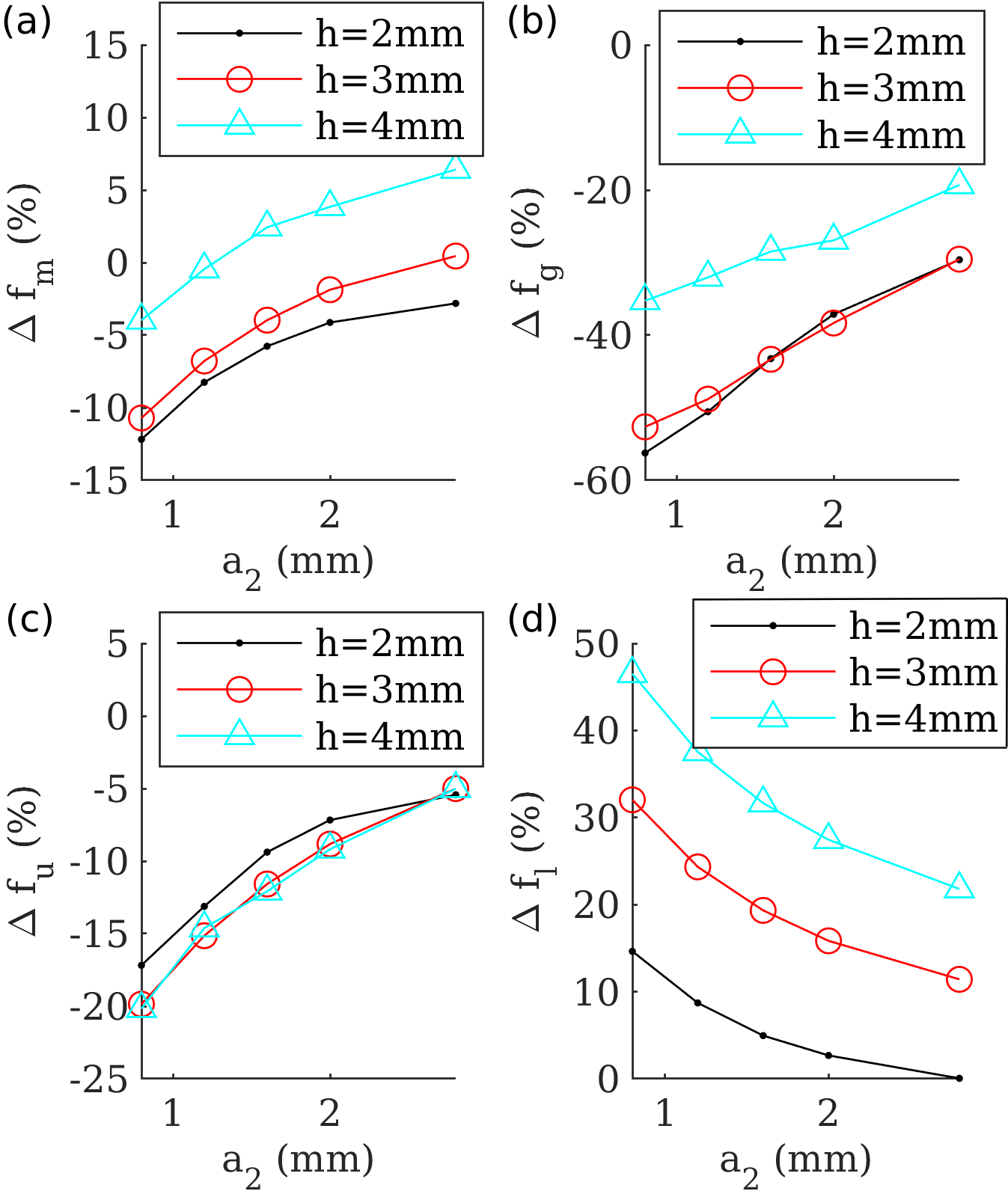} 
 \label{fig:band_diag_parameters}
 \caption{\label{fig:band_diag_parameters}: Calculated percent difference values between analytical and FEA results for uncovered cell width values $a_2\in \{0.8,...,4.0\}$ and height values $h=2$ mm, $h=3$ mm, and $h=4$ mm. Percent difference in (a) the middle of the band gap $\Delta f_\mathrm{m}$, and (b) the width of the band gap $\Delta f_\mathrm{g}$. Percent difference in (c) the eigenfrequency at $k=0$ ($\Delta f_\mathrm{u}$), and (d) the eigenfrequency at $k=1$ ($\Delta f_\mathrm{l}$).}
\end{figure}

While the analytical method closely agrees with the FEA calculated $f_\mathrm{m}$, the band gap width $f_\mathrm{g}$, and the eigenfrequencies $f_\mathrm{u}$, $f_\mathrm{l}$ show greater disagreement. The analytical method underestimates $f_\mathrm{u}$ by $5$-$20\%$. The analytical method overestimates the eigenfrequency of the $S_0$ mode at $k=1$ ($f_\mathrm{l}$) by $0.1$-$47\%$. For lower values of $h$, and higher values of $a_2$, $|\Delta f_\mathrm{l}|$ and $|\Delta f_\mathrm{u}|$ are reduced. Figure \ref{fig:band_diag_parameters} (b) shows how the approximate analytical method underestimates the width of the band gap by $19$-$56\%$. This underestimation also reduces with an increase in $a_2$, but reduces with a larger $h$. These calculations demonstrate that the agreement between the analytical method and the FEA method depends greatly on the thickness $h$ and electrode gap $a_2$. 

To investigate the effect of the electrical capacitance equations on calculated band diagrams, calculated capacitances from FEA steady state electrostatic simulations were compared with capacitances calculated using equations (\ref{eq:C1})--(\ref{eq:C3}). Figure \ref{fig:mult_factors} (a)--(c) shows plots of the capacitance values for a set of width values $a_2\in \{0.2,...,4.0\}$.

\begin{figure}[!ht]
 \centering
 	\includegraphics[scale=1]{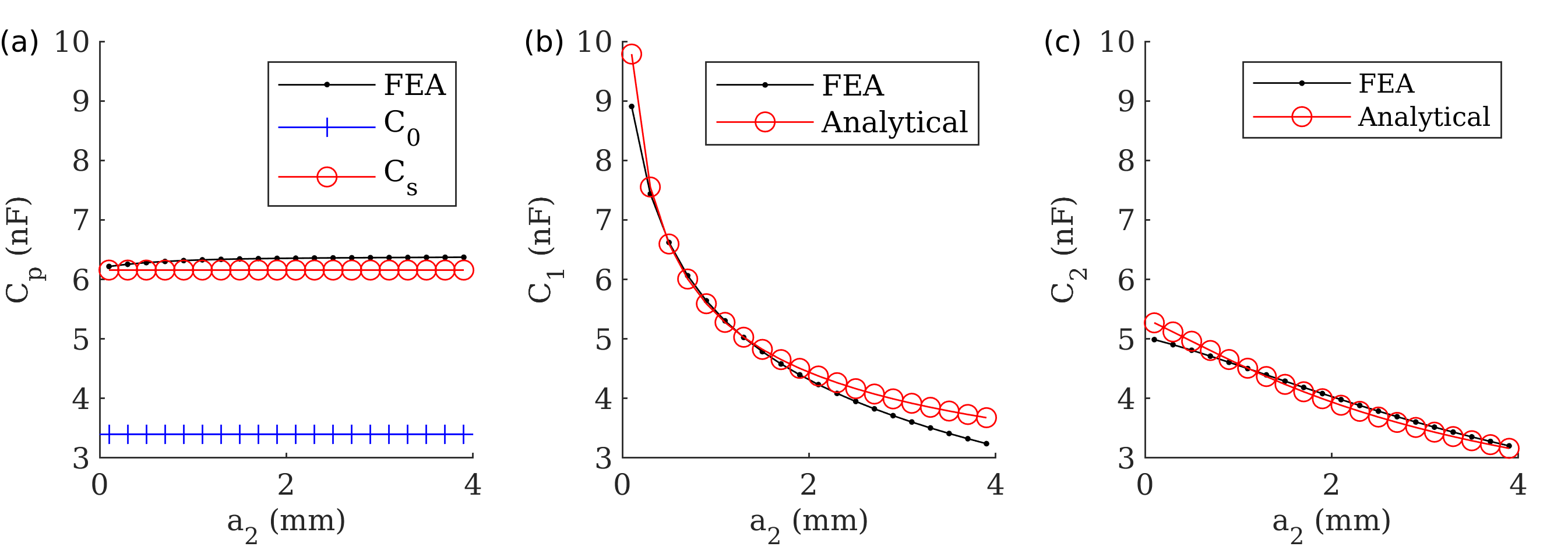} 
 \label{fig:mult_factors}
 \caption{\label{fig:mult_factors}: Mutual capacitance calculated by FEA and approximate analytical methods for uncovered cell width values $a_2\in \{0.2,...,4.0\}$. (a) Capacitance between parallel plates $C_\mathrm{p}$, calculated using Eq. (\ref{eq:Cp}), (b) capacitance between co-planar nearest-neighbor electrodes $C_1$, calculated using Eq. (\ref{eq:C1}), and (c) capacitance between off-center parallel plates $C_2$, calculated using Eq. (\ref{eq:C3}).}
\end{figure}

Figure (\ref{fig:mult_factors}) shows a close agreement in capacitance values between the analytical methods and the FEA simulation. In Figure (\ref{fig:mult_factors}) (a) it can be seen that the capacitance between parallel plates for this geometry, calculated using Eq. (\ref{eq:Cp}) to include fringing effects is nearly double the capacitance calculated using the simple parallel plate equation (\ref{eq:C0}). The capacitance $C_\mathrm{p}$ is within 3.4\% of the capacitance values from the simulation for all values of $a_2$. The top electrode nearest neighbor capacitance $C_1$ is within 13\% of the capacitance values from the simulation for all values of $a_2$. The greater discrepency in $C_1$ for $a_2>2$mm is due to the limitation of the method for calculated capacitance when the distance between electrodes exceeds the thickness of the material. The capacitance between top and nearest neighbor bottom electrodes $C_3$ agreed well with simulated values, staying within 5.7\% of the FEA values for all values of $a_2$. Based on how well the FEA calculated capacitances compare with the analytically calculated capacitances, it is likely that innacuracies in computed capacitances contribute to some, but not all, of the band diagram discrepencies between analytical and FEA methods.

\subsection{Scattering parameters for finite plates}

Finite plate models with $N=6$ electrodes were used to compute global scattering coefficients with Equation (\ref{eq:global_scattering}). The example geometry of $h=2$ mm, $a_1=0.5$ mm, $a_2=0.5$ mm was used with $L=1$ mH to avoid Fabry Perot resonances in the plate and to ensure $\lambda\ll\Lambda$ near the hybridization band gap. The reflection coefficient $S^\mathrm{G}_{11}$ and the transmission coefficient $S^\mathrm{G}_{13}$ for the acoustic wave is shown for frequencies $f\in\{0.01,0.1\}$ MHz in Figure \ref{fig:scattering_coefficients} (a). The reflection coefficient $S^\mathrm{G}_{22}$ and the transmission coefficient $S^\mathrm{G}_{24}$ for the electric potential wave is shown in Figure \ref{fig:scattering_coefficients} (b) below. While the computed global reflection coefficients change with the number of unit-cells $N$, Figure \ref{fig:scattering_coefficients} gives an example of what the scattering coefficient look like for band gaps near $\lambda\ll\Lambda$ with a practical amount of electrode pairs in the array. 

 \begin{figure}[!ht]
\includegraphics[scale=1]{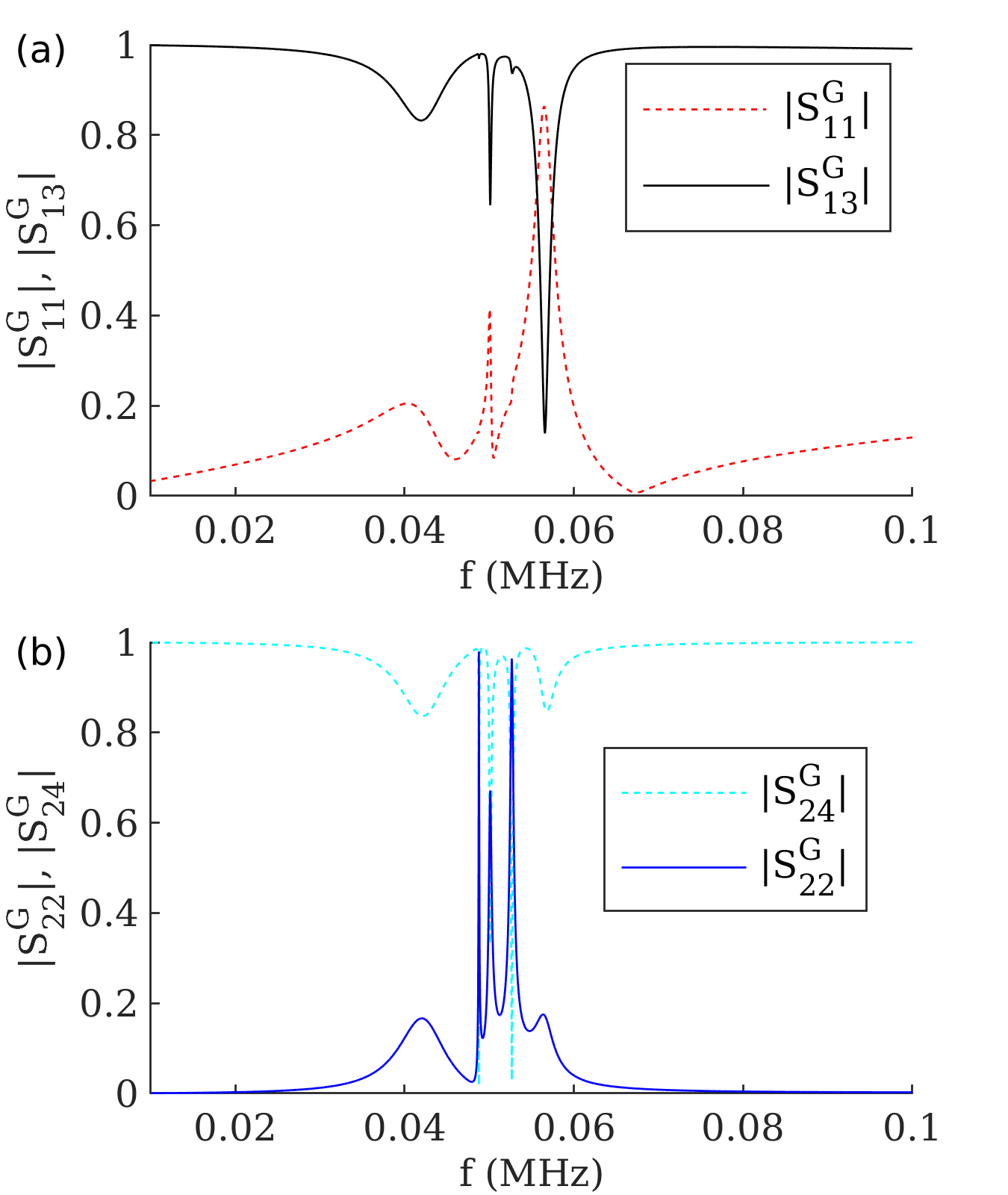} 
 \label{fig:scattering_coefficients}
 \caption{\label{fig:scattering_coefficients}: (a) Global reflection and transmission coefficients $S^\mathrm{G}_{11}$ and $S^\mathrm{G}_{13}$ for the elastic wave. (a) Global reflection and transmission coefficients $S^\mathrm{G}_{22}$ and $S^\mathrm{G}_{24}$ for the electrical wave.}
\end{figure}

The scattering coefficients in Figure \ref{fig:scattering_coefficients} have multiple inflection frequencies, associated with different propagation modes. Near 42 kHz, there is a decrease in $S^\mathrm{G}_{24}$ and $S^\mathrm{G}_{13}$. This is associated with the first mode of propagation, shown symbolically by the mutual capacitances in Figure \ref{fig:cap_diag} (a). The decrease in transmission at 57 kHz is associated with the second mode of propagation, which is shown symbolically by the mutual capacitances in Figure \ref{fig:cap_diag} (b). For this geometry, this occurs at a slightly higher frequency than the first mode, and is more pronounced for a $S_0$ mode excitation. 

To show how the scattering parameters change with varying the inductance $L$ and by varying the height $h$, the middle of the second mode inflection corresponding to a band gap $f_\mathrm{m}$ was calculated, as well as the quality factor $Q$. The quality factor is defined as $Q=f_\mathrm{m}/(f_{\mathrm{w1}}-f_{\mathrm{w2}})$, where $f_{\mathrm{w1}}$ and $f_{\mathrm{w2}}$ are the first and second frequencies at half of the decrease in transmission $S^\mathrm{G}_{13}$. Figure \ref{fig:effective_parameters} (a) shows $f_\mathrm{m}$ and $Q$ for $a_1=0.5$ mm, $a_2=0.5$ mm, $L=1$ mH, and $h\in\{2$,$3$,$4\}$ mm, while Figure \ref{fig:effective_parameters} (b) shows $f_\mathrm{m}$ and $Q$ for $a_1=0.5$ mm, $a_2=0.5$ mm, $h=2$ mm, and $L\in\{1,10,100,1000\}$ mH.

 \begin{figure}[!ht]
\includegraphics[scale=1]{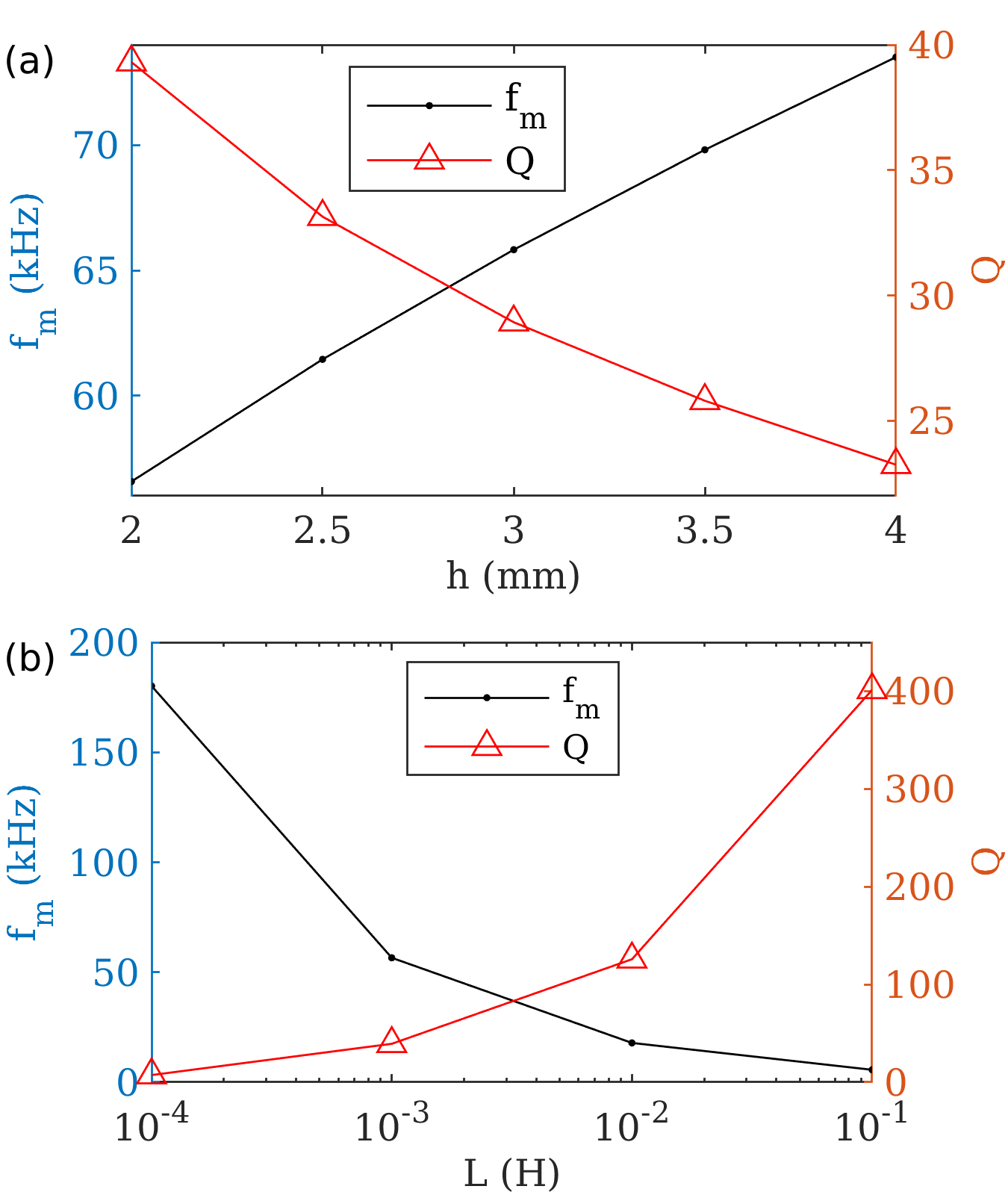} 
 \label{fig:effective_parameters}
 \caption{\label{fig:parameter_vary}: The middle of the inflection in $S^\mathrm{G}_{13}$ corresponding to the hybridization band gap $f_\mathrm{m}$, and the quality factor $Q$ for (a) $a_1=0.5$ mm, $a_2=0.5$ mm, $L=1$ mH, and $h\in\{2$,$3$,$4\}$ mm and (b) $a_1=0.5$ mm, $a_2=0.5$ mm, $h=2$ mm, and $L\in\{1,10,100,1000\}$ mH.}
\end{figure}

The band gap moves to higher frequencies with an increase in $h$. This is shown in Figure \ref{fig:parameter_vary} (a), with $f_\mathrm{m}=56.6$kHz for $h=2$mm, and $f_\mathrm{m}=73.5$kHz for $h=4$mm. The quality factor $Q$, however, gets smaller with an increase in $h$. This is to be expected, due to the electrical boundary conditions having a less pronounced effect on the scattering parameters as $h$ is increased. 

The band gap moves to lower frequencies with an increase in inductance $L$. This is shown in Figure \ref{fig:parameter_vary} (b), with $f_\mathrm{m}=180$kHz for $L=10^{-4}$ H, and $f_\mathrm{m}=5.65$kHz for $L=10^{-1}$mm. The quality factor $Q$ also increases with an increase in $L$. Therefore, if high $Q$ is to be desired for an application, increasing $L$ until the desired band gap frequency is reached would be the simplest design consideration.
 
The frequency shift in $f_\mathrm{m}$ due to inductance is expected and predicted by an infinite circuit model of this problem. If we consider the infinite plate model, the hybridization band gap is an avoided crossing with the purely electrical circuit band and the $S_0$ mode lamb wave band. A simplified electrical circuit for the first electrical mode (Figure \ref{fig:cap_diag} (a)) used in \cite{kherraz2018hybridization,kherraz2019tunable} is

\begin{equation}
\sin( \Lambda/2)=\sqrt{\frac{1}{LC_1\omega^2}-\frac{C_\mathrm{p}}{C_1}}
\label{eq:simple_electrical_dispersion}
\end{equation}

From this equation, it can be predicted that the electrical band will shift to lower frequencies with an increase in $L$ or $C_1$. The upper frequency of the band gap is predicted in the above equation by the LC resonance, which for $h=2$ mm, $a_1=1$ mm, and $a_2=2$ mm is $1/\sqrt{L C_0}=203$ kHz. The lower frequency is predicted to be $95.4$ kHz. The approximate analytical method presented in this article gives $f^\mathrm{A}_\mathrm{u}=157$ kHz and $f^\mathrm{A}_\mathrm{l}=122$ kHz, and the FEA gives $f^\mathrm{F}_\mathrm{u}=170$ kHz and $f^\mathrm{F}_\mathrm{l}=119$ kHz. Though the above equation (\ref{eq:simple_electrical_dispersion}) gives larger differences in band gap estimation with FEA than the analyical method presented in this article, it may be useful for building intuition on the affect of $L$ and geometry that changes $C_1$ and $C_0$, and for estimating the middle of the band gap $f_\mathrm{m}$.

\subsection{Calculation of Effective Parameters}

The effective elastic impedance and effective elastic wavenumber were calculated using the global scattering parameters. Equation  (\ref{eq:equivalent_wavenumber}) was used to calculate the complex effective wavenumber, which is plotted vs. frequency in Figure \ref{fig:effective_parameters} (a). The branch $q$ was chosen a each $\omega$ to reduce wrapping jumps. Equation (\ref{eq:equivalent_impedance}) was used to calculate the effective impedance, the absolute value of which is plotted vs. frequency in Figure (\ref{fig:effective_parameters}) (b). 

 \begin{figure}[!ht]
\includegraphics[scale=1]{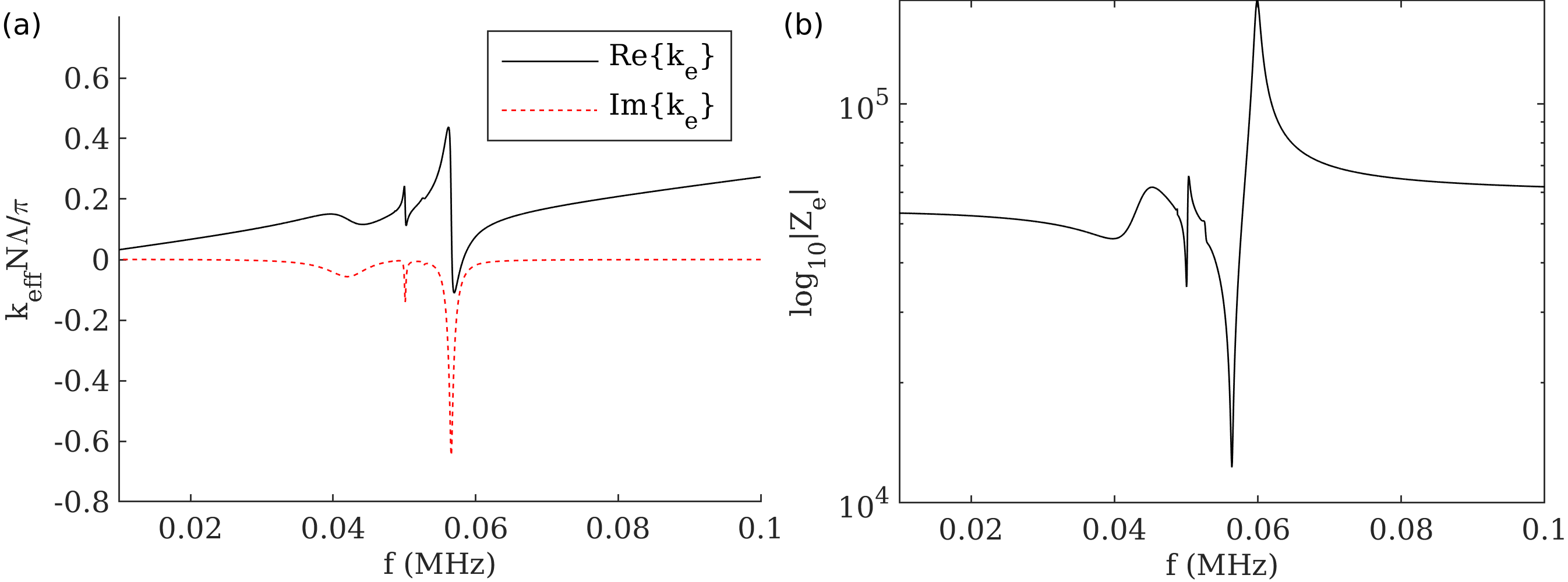} 
 \label{fig:effective_parameters}
 \caption{\label{fig:effective_parameters}: Effective elastic parameters for frequencies $f\in{0.01,0.1}$ MHz using a finite plate of $N=6$ electrode pairs, $h=2$ mm, $a_1=0.5$ mm, $a_2=0.5$ mm, and $L=1$ mH. (a) Effective real and imaginary parts of the wavenumber ($Re\{k_{\mathrm{eff}}\}$ and $Im\{k_{\mathrm{eff}}\}$), (b) absolute value of the effective impedance $|Z_\mathrm{e}|$, plotted on a log-scale.}
\end{figure}

The effective impedance and wavenumber, plotted in Figure \ref{fig:effective_parameters}(b) and (c), show local minimum and maximum values that correspond to the inflections of the scattering parameters $S^\mathrm{G}_{11}$ and $S^\mathrm{G}_{31}$ plotted in Figure \ref{fig:effective_parameters}(a). The minima of the effective impedance at 57 kHz coincides with the normalized effective wavenumber dropping from +0.44 to -0.11. The relatively smaller jumps in effective wavenumber at lower frequencies also correspond to inflections in the scattering parameters.

A finite plate length with $N=6$ yields a different wavenumber calculation than an infinitely periodic plate. In this example, if $N=10$, the effective wavenumber drops from +0.64 to -0.16 at the slightly higher frequency 58 kHz. It was found that if the number of electrode pairs, or unit cells, is increased, the change in effective wavenumber at the band gap is increased and the frequency where the change occurs slightly increases. This is likely do to edge effects of a finite 2D system. More unit cells will allow the system parameters to converge to parameters for the infinite lattice case. For the extraction of the infinite lattice wavenumber from a finite lattice, one can use a procedure similar to the one detaled in \cite{roux2020homogenization}.

As can be observed in Fig. 10(a), the effective wavenumber exhibits extreme characteristics in and around the hybridization band gap, including large, near-zero and and negative values.  These values of wavenumber, while similar to those found at higher phononic band gaps, are achieved at significantly lower frequencies within the homogenization limit that are typically only found in deeply subwavelength elements in acoustic or elastic metamaterials, such as transmission-line arrangements of Helmholtz resonators \cite{fang2006ultrasonic}.  Unlike passive metamaterial effective properties, however, the use of a piezoelectric material with shunted circuits allows for electrical tuning of the hybridization bandgap, and therefore offer the desireable ability for electrical tuning of the elastic wave propagation characteristics in these piezoelectric metamaterials.

\section{Conclusions}
A 2D approximate analytical method for modeling piezoelectric plates with periodic shunted electrodes has been presented. This method formulates the governing piezoelectric equations for symmetric Lamb wave propagation into a convenient impedance matrix for homogeneous sections of plate. The impedance matrix can be converted into a transfer matrix or a scattering matrix, and can be coupled with other matrices to model plates with non-homogeneous cross-sections, properties, and electrical boundary conditions.

A transfer matrix method with scattering parameter matrices was used to calculate band diagrams and global reflection and transmission coefficients for a homogeneous cross-section, uniformly poled plate with non-homogeneous electrical boundary conditions. Band diagrams constructed using the approximate analytical method were compared to ones calculated using FEA. The band diagrams compared well, especially at frequencies much lower than the Bloch wave band gap. The estimation of the low-frequency band gap resulting from electro-mechanical coupling in the plate was used as the metric for comparing the two methods. When compared to the FEA results, the analytical method better estimated the low-frequency band gap for greater distances between electrodes $a_2$. It was found that lower ratios $\mu=a_1/a_2$ and higher ratios of $a_1/h$ contributed to better comparisons. 

The deviations in band gap estimation seem largely due to the neglected elecro-mechanical coupling in sections of the plate that are not covered by electrodes, and due to the use of lumped impedances to model the electrical conduction in those uncovered sections. The analytically calculated capacitances ($C_\mathrm{p}$,$C_1$, and $C_2$) used to model electrical conduction in uncovered sections compared well with FEA calculated capacitances. Still, experiments are needed to test how effective the approximate methods are in predicting capacitances, dispersion relations, and scattering coefficients.

The analytical method allowed the calculation of effective impedance and effective wavenumber for plates of varying lengths. An example was given that showed the effective wavenumber around the hybridization band gap drop to negative real and imaginary values after climbing to relatively large positive real values. The impedance loads in the periodically spaced shunted circuits are the electro-mechanical  analog to a waveguide lined with Helmholtz resonators.

The presented approximate analytical method gives researchers a workable model for scattering parameters at a time when electrical tuning and additive manufacturing are changing the possible designs and applications of piezoelectric metamaterials. The convenient formulation of the relevant equations into an impedance matrix gives researchers the opportunity to model space-time modulated piezoelectric metamaterials in the frequency domain using a transfer matrix method, as was done with Helmholtz resonator acoustic metamaterials \cite{li2019transfer,shen2019nonreciprocal}. Further, this formulation will allow researchers to model curved and stepped cross-sectional geometries that are now more practical to manufacture due to concurrent research in additive manufacturing \cite{schipf2022barium,cui2019three}. 

\section{acknowledgments}
This work was supported by the US Office of Naval Research and a National Research Council (NRC) Postdoctoral Fellowship award at the US Naval Research Laboratory. 

\section{Appendix A}
\label{sec:AppendixA}
The scattering matrix for a cell relates the amplitudes of the waves on both sides. The amplitudes of waves, represented by vectors $\mathbf{\psi}^{'}$ have frontward and backward components, represented with superscripts $+$ and $-$, respectively. With port 1 on the left side of a cell, and port 2 on the right, the relation between wave amplitudes for cell $A$ can be given as 
\begin{equation}
\begin{pmatrix}
\mathbf{\psi}^{'-}_1\\
\mathbf{\psi}^{'+}_2\\
\end{pmatrix}=
\begin{pmatrix}
\mathbf{S}^\mathrm{A}_{11} & \mathbf{S}^\mathrm{A}_{12}\\
\mathbf{S}^\mathrm{A}_{21} & \mathbf{S}^\mathrm{A}_{22}\\
\end{pmatrix}
\begin{pmatrix}
\psi^{'+}_1\\
\psi^{'-}_2\\
\end{pmatrix}
\end{equation}
where subscript $'$ denotes the wave state on the outside of the cell boundary, and subscript $+$ and $-$ denote frontward and backward waves, respectively. The scattering parameters $S^\mathrm{A}_{ij}$, for a 4 port cell are 2x2 block matrices, and $\mathbf{\psi}^{'}_i$ are 2x1 vectors. The same scattering matrix equation can be used for the neighboring cell, which we label as cell $B$. 

The Redheffer star product can be used to construct a scattering matrix of two cells from the scattering matrices of each of the cells. The Redheffer star product between the scattering matrices of cells $\mathrm{A}$ and cell $\mathrm{B}$ is \cite{rumpf2011improved}

\begin{equation}
\mathbf{S}^\mathrm{A}\otimes \mathbf{S}^\mathrm{B} = \begin{pmatrix}
\mathbf{S}^\mathrm{A}_{11}+\mathbf{S}^\mathrm{A}_{12}(\mathbf{I}-\mathbf{S}^\mathrm{B}_{11}\mathbf{S}^\mathrm{A}_{22})^{-1}\mathbf{S}^\mathrm{B}_{11}\mathbf{S}^\mathrm{A}_{21} & \mathbf{S}^\mathrm{A}_{12}(\mathbf{I}-\mathbf{S}^\mathrm{B}_{11}\mathbf{S}^\mathrm{A}_{22})^{-1}\mathbf{S}^\mathrm{B}_{12}\\
\mathbf{S}^\mathrm{B}_{21}(\mathbf{I}-\mathbf{S}^\mathrm{A}_{22}\mathbf{S}^\mathrm{B}_{11})^{-1}\mathbf{S}^\mathrm{A}_{21} & \mathbf{S}^\mathrm{B}_{22}+\mathbf{S}^\mathrm{B}_{21}(\mathbf{I}-\mathbf{S}^\mathrm{A}_{22}\mathbf{S}^\mathrm{B}_{11})^{-1}\mathbf{S}^\mathrm{A}_{22}\mathbf{S}^\mathrm{B}_{12}\\
\end{pmatrix} .
\end{equation}
The Redheffer star product can be used to find a global scattering matrix for a group of cells, or to find the dispersion relation for an infinitely periodic plate waveguide.

%

\end{document}